\documentclass[fp,twocolumn]{jpsj3}

\usepackage{txfonts}
\usepackage[dvipdfmx]{graphicx}
\usepackage{bm}
\usepackage{color}
\usepackage{amsmath,amssymb}
\usepackage{array}
\usepackage{epstopdf}
\title{Magnetoelectric Response in Electric Octupole State: \\Possible Hidden Order in Cuprate Superconductors}

\author{Takanori Hitomi and Youichi Yanase\thanks{yanase@scphys.kyoto-u.ac.jp}}
\inst{Department of Physics, Graduate School of Science, Kyoto University, Kyoto 606-8502, Japan}

\date{December 19, 2018}

\abst{
Motivated by recent studies of odd-parity multipole order in condensed matter physics, 
we theoretically study magnetoelectric responses in an electric octupole state.
Investigating the Edelstein effect and spin Hall effect in a locally noncentrosymmetric bilayer Rashba model,
we clarify characteristic properties due to parity violation in the electric octupole state. 
Furthermore, a possible realization of electric octupole order in bilayer high-$T_c$ cuprate superconductors is proposed. 
Our calculation of magnetic torque is consistent with recent experimental observation of a kink above 
the superconducting transition temperature. 
We also show significant enhancement of the in-plane anisotropy in spin susceptibility due to the superconductivity, 
and propose an experimental test by means of the nuclear magnetic resonance in the superconducting state. 
A spin-orbit coupled metal state in Cd$_2$Re$_2$O$_7$ is also discussed. 
}

\begin{document}
\maketitle

\section{Introduction}

Advances in research of noncentrosymmetric electron systems have revealed a variety of intriguing phenomena, 
such as exotic or topological superconductivity~\cite{Edel'shtein_SC_JETP1989,Edel'shtein_ME_PRL1995,Gor'kov-Rashba,Frigeri_SC,Kaur_SC,Yip_SC_ME_PRB2002,Fujimoto_SC_ME_PRB2005,Fujimoto_SC_ME_JPSJ2007_1,
Fujimoto_SC_ME_JPSJ2007_2,Yanase2007,NCSCbook,Sato2009,Sau2010,Alicea2010,Lutchyn2010,Mourik2012,
Nadj-Perge2014,Yanase2008,Daido2016,Takasan_TS}, magnetoelectric effect~\cite{
Edel'shtein_ME_PRL1995,Levitov_ME_JETP1985,Yip_SC_ME_PRB2002,Fujimoto_SC_ME_PRB2005,Fujimoto_SC_ME_JPSJ2007_1,Fujimoto_SC_ME_JPSJ2007_2,Edelstein1990}, anomalous Hall effect~\cite{Fujimoto_SC_ME_JPSJ2007_1,Fujimoto_SC_ME_JPSJ2007_2}, spin Hall effect~\cite{Murakami_SHE,Sinova_SHE,Inoue_SHE,Kontani_SHE,Mizoguchi_SHE}, and topological transport~\cite{Kane_QSHE1,Kane_QSHE2,Bernevig_QSHE}.
Interplay of relativistic spin-orbit coupling and broken inversion symmetry gives rise to antisymmetric spin-orbit coupling (ASOC) such as Rashba spin-orbit coupling, which results in these phenomena due to spin-momentum locking in electronic band structures.

ASOC entangles not only spin, orbital, and momentum but also sublattice degrees of freedom 
in locally noncentrosymmetric systems, 
where the crystal structure preserves global inversion symmetry although local site symmetry lacks inversion symmetry. 
Such crystals have at least two nonequivalent lattice sites, between which ASOC changes the sign. 
As a consequence of the sublattice-dependent ASOC, ``spin-momentum-sublattice locking'' occurs, 
which is an analog of spin-momentum locking in globally noncentrosymmetric systems. 
The peculiar electronic structure has been studied for superconductivity~\cite{Fischer2011,Maruyama_SC,Sigrist_SC_NCS}, 
and later demonstrated by experiments~\cite{Goh2012,Shimozawa2016} and first principles band structure 
calculations~\cite{Zhang2014-2,Riley2014,Jones2014,Ghelmann2016,Klein2016} for various locally noncentrosymmetric 
crystalline compounds.

Recent theoretical works have shown exotic superconductivity~\cite{Nakosai2012,Yoshida_PDW_PRB2012,Yoshida2013,Yoshida_PDW_JPSJ2014,Watanabe_PDW_PRB2015,Yoshida_TS_PRL2015,Higashi2016,Yoshida_Tsuneya2017,Sigrist_SC_NCS} and odd-parity multipole order~\cite{Yanase_ME,Sumita_MQ,Hayami_TO_PRB2014,Hayami_TO_PRB(R)2014,Hayami_TO_JPSJ2015,Sugita_TI,
Hitomi_EO_JPSJ2014,Hitomi_EO_JPSJ2016,Watanabe_BaMn2As2,Hayami2017,Yanagi2017,Hayami2017b,Watanabe2018a,Hayami2018,Hayami2018b} 
induced by the sublattice-dependent ASOC. 
The former may be a platform of topological superconductivity~\cite{Nakosai2012,Watanabe_PDW_PRB2015,Yoshida_TS_PRL2015,Yoshida_Tsuneya2017}, and the latter is parity-violating electromagnetic order. 
Experimental study traces back to magnetic monopole order in Cr$_2$O$_3$~\cite{Cr2O3}, and recent studies found
magnetic toroidal order in LiCoPO$_4$~\cite{Aken_LiCoPO4,Spaldin_LiCoPO4}, 
magnetic quadrupole order in Sr$_2$IrO$_4$~\cite{Zhao_Sr2IrO4,Matteo_Sr2IrO4} and Ba(TiO)Cu$_4$(PO$_4$)$_4$~\cite{Kimura_Ba(TiO)Cu4(PO4)4}, 
and odd-parity electric order in Cd$_2$Re$_2$O$_7$~\cite{Hiroi_Cd2Re2O7,Harter2017}. 
Recent theoretical study has also identified magnetic hexadecapole order in BaMn$_2$As$_2$.~\cite{Watanabe_BaMn2As2} 
Furthermore, a comprehensive group theoretical classification has identified more than 100 compounds 
as odd-parity magnetic multipole states~\cite{Watanabe2018a}. 
These multipole states spontaneously break global inversion symmetry.
The combined effects of spontaneous parity violation and sublattice-dependent ASOC may lead to 
fascinating magnetic, superconducting, and transport phenomena. 
For instance,  Fulde-Ferrell-Larkin-Ovchinnikov superconductivity by multipole order has been suggested~\cite{Sumita_MQ}, 
and possible realization in Sr$_2$IrO$_4$ was proposed~\cite{Sumita_Sr2IrO4}.

In this paper, we study magnetic and magnetoelectric transport properties in the electric octupole (EO) state of bilayer Rashba systems, which was previously proposed for Sr$_3$Ru$_2$O$_7$~\cite{Hitomi_EO_JPSJ2014,Hitomi_EO_JPSJ2016}. 
Owing to the layer degrees of freedom, the EO state is realized by antiferroic stacking of the even-parity electric quadrupole (EQ) moment between bilayers.
Spin texture characteristic of the EO state appears in the momentum space.~\cite{Hitomi_EO_JPSJ2014}
Therefore, intriguing magnetoelectric responses may arise from hidden spin polarization in the EO state.

We discuss high-temperature cuprate superconductors as a possible candidate.
Recently, a magnetic-torque measurement explored second-order phase transition around the pseudogap 
onset temperature in a bilayer cuprate superconductor YBa$_{2}$Cu$_{3}$O$_{7-x}$ (YBCO)~\cite{Sato_YBCO}.
A kink in the magnetic torque indicates a change in the nematicity at the critical point. 
However, nematic order breaking the $C_4$ rotation symmetry can not be a primary order parameter 
since the $C_4$ rotation symmetry is already broken in the orthorhombic YBCO owing to CuO chains. 
Indeed, the critical exponent observed in experiments~\cite{Sato_YBCO} indicates that the nematicity is not a primary order parameter. 
On the other hand, optical measurements such as linear dichroism~\cite{Lubashevsky_optical} 
and second-harmonic optical anisotropy~\cite{Zhao2016} measurements have shown 
broken inversion symmetry. Thus, the hidden order in cuprates may be attributed to the EO order which is closely related to the nematicity and breaks inversion symmetry. In this paper, consistency with magnetic torque measurements~\cite{Sato_YBCO} is demonstrated. Furthermore, the spin Hall effect and Edelstein effect are calculated, and their characteristic properties are proposed for future experimental tests.

This paper is constructed as follows.
First, in Sec.~2.1 we introduce the forward scattering model for bilayer Rashba systems and formulate 
a mean field theory. Next, we introduce EQ and EO order in Sec.~2.2. 
In Sec.~3, we study the EO state in an orthorhombic system and discuss the recent experimental result in YBCO. 
In Sec.~3.1 it is shown that the nematicity changes at the phase transition temperature of EO order.
As a consequence, a kink appears in magnetic torque in agreement with the experiment~\cite{Sato_YBCO}  (Sec.~3.2).
Furthermore, in Sec.~3.3 we show that in-plane anisotropy in magnetic susceptibility is remarkably enhanced in the superconducting state. 
Transport properties characterizing the EO state are elucidated in Secs.~4 and 5. 
We calculate the spin Hall conductivity in Sec.~4 and the Edelstein effect in Sec.~5. 
In addition to high-$T_{\rm c}$ cuprate superconductors, the spin-orbit coupled metallic state in Cd$_2$Re$_2$O$_7$ is briefly discussed.
Finally, summary and discussions are given in Sec. 6.

\section{Model}
\subsection{Bilayer Rashba model}
In order to investigate odd-parity EO order in bilayer Rashba systems,
we introduce a forward scattering model~\cite{Hitomi_EO_JPSJ2016},
\begin{align}
  H &= H_{\rm{kin}} + H_{\rm{ASOC}} + H_{\perp} + H_{\rm{f}} ,   \label{H_all} \\
  H_{\rm{kin}}   &= \sum_{\bm{k}} \sum_{s=\uparrow, \downarrow} \sum_{l=A,B} \varepsilon_{\bm{k}} c^{\dagger}_{\bm{k}sl} c_{\bm{k}sl} ,  \label{H_kin} \\
  H_{\rm{ASOC}}  &= \sum_{\bm{k},s,s',l} \alpha_{l} \bm{g}_{\bm{k}} \cdot \bm{\sigma}^{ss'} c^{\dagger}_{\bm{k}sl} c_{\bm{k}s'l} ,   \label{H_ASOC} \\
  H_{\perp}     &= t_{\perp} \sum_{\bm{k},s} [c^{\dagger}_{\bm{k}sA} c_{\bm{k}sB} + \rm{h.c.}] ,           \label{H_perp} \\ 
  H_{\rm{f}}    &= - \frac{g_{1}}{2N} \sum_{\bm{k},\bm{k}',l} d_{\bm{k}} d_{\bm{k}'} n_{\bm{k}l} n_{\bm{k}'l} \notag \\
  & \hspace{0mm} - \frac{g_{2}}{2N} \sum_{\bm{k},\bm{k}'} d_{\bm{k}} d_{\bm{k}'} [n_{\bm{k}A} n_{\bm{k}'B} + n_{\bm{k}B} n_{\bm{k}'A}], \label{H_f}
\end{align}
where $c_{\bm{k}sl}$ ($c^{\dagger}_{\bm{k}sl}$) is an annihilation (creation) operator
of an electron with spin $s=\uparrow,\downarrow$ and wave vector $\bm{k}$ on a layer $l=A,B$.
$n_{\bm{k}l} = \sum_{s} c^{\dagger}_{\bm{k}sl} c_{\bm{k}sl}$ is number density operator,
$\bm{\sigma} = (\sigma_{x}, \sigma_{y}, \sigma_{z})$ is the Pauli matrices,
and $N$ is the number of sites per layer.
In the kinetic energy term $H_{\rm{kin}}$, $\varepsilon_{\bm{k}}$ is represented
by the nearest- and next-nearest-neighbour hoppings in a two-dimensional (2D) square lattice,
$\varepsilon_{\bm{k}} = -2 t_{x} \cos k_{x} -2 t_{y} \cos k_{y} - 4t_{2} \cos k_{x} \cos k_{y} - \mu$,
where the chemical potential $\mu$ is included.
The nearest neighbour hoppings are given by
\begin{align}
  t_{x} &= t_{1} + \delta t_{1} , \label{tx} \\
  t_{y} &= t_{1} - \delta t_{1}   \label{ty},
\end{align}
where $\delta t_{1}$ introduces an orthorhombic anisotropy in the 2D plane.
Thus, $\delta t_{1} = 0$ in the tetragonal system while $\delta t_{1} \neq 0$ in the orthorhombic system such as YBCO.
We adopt $t_{1} = 1$ as the unit of energy. 

The second term $H_{\rm{ASOC}}$ stands for layer-dependent Rashba ASOC.
Since the parity-violating crystalline electric field is opposite between the two layers,
the coupling constant has the form $(\alpha_{A} ,\alpha_{B}) = (\alpha , -\alpha)$. 
This layer-dependent Rashba ASOC has been observed in recent spin- and angle-resolved photo-emission 
spectroscopy for a bilayer cuprate superconductor Bi2212,~\cite{Gotlieb2018} 
and a sizable magnitude $\alpha \sim 10$meV was reported. 
In accordance with Ref.~\citen{Gotlieb2018}, We assume a simple form $\bm{g}_{\bm{k}} = (-\sin k_{y} ,\sin k_{x}, 0)$, which has been microscopically derived 
when the orbital degrees of freedom are quenched~\cite{Yanase_Sr2RuO4}. 
The third term $H_{\perp}$ represents the interlayer hopping of electrons.
Assuming quasi-2D bilayer systems, a small interlayer hopping amplitude is taken as $t_{\perp} = 0.1$.  

The last term $H_{\rm{f}}$ in Eq.~(\ref{H_all}) describes intralayer and interlayer forward scattering interactions, 
whose coupling constants are $g_{1}$ and $g_{2}$, respectively. 
A $d$-wave form factor $d_{\bm{k}} = \cos k_{x} - \cos k_{y}$ is adopted. 
The intralayer term has been obtained as an effective interaction by the renormalization group theory
for the 2D Hubbard model~\cite{Halboth_RG_PRL2000,Halboth_RG_PRB2000,Honerkamp_RG,Metzner_RG}. 
It leads to spontaneous deformation of the Fermi surface,~\cite{Yamase_tJ,Khavkine_PI,Kee_PI,Yamase_PI_1} which is 
called $d$-wave Pomeranchuk instability (dPI).
The dPI arising from the spin-fluctuation was also revealed on the basis of the $d$-$p$ model~\cite{Kawaguchi_CDW}.   
The dPI is equivalent to EQ order from the viewpoint of symmetry.
Indeed, the local EQ moment with $O_{x^2-y^2}$ symmetry appears at each layer. 
When the stacking of EQ moment is antiferroic (ferroic) between layers, the system undergoes EO order (EQ order). 
Although the sign of interlayer interaction plays an essential role for the stability of these states,~\cite{Hitomi_EO_JPSJ2016} derivation from the bilayer Hubbard or $d$-$p$ model has not been carried out. 
Thus, we phenomenologically introduce a negative (positive) interlayer forward scattering term in order to examine the EO (EQ) state. Consistency with experimental results is obtained only for the EO state. We leave the derivation of the $g_2$ term as a future issue.

We treat the forward scattering interaction term $H_{\rm f}$ by a mean field approximation.
By decoupling $n_{\bm{k} l} n_{\bm{k}' l'} \simeq  n_{\bm{k} l} \langle n_{\bm{k}' l'} \rangle + \langle n_{\bm{k}l} \rangle n_{\bm{k}' l'} - \langle n_{\bm{k} l} \rangle \langle n_{\bm{k}' l'} \rangle $,
the dPI order parameter on the $l$-layer is introduced as
\begin{equation}
  \Delta_{l} = \Delta_{1 l} + \Delta_{2 \bar{l}} ,   \label{OP_l}
\end{equation}
with 
\begin{align}
  \Delta_{1 l} &= - \frac{g_{1}}{N} \sum_{\bm{k}} d_{\bm{k}} \langle n_{\bm{k}l} \rangle,    \label{OP_l_intra} \\
  \Delta_{2 l} &= - \frac{g_{2}}{N} \sum_{\bm{k}} d_{\bm{k}} \langle n_{\bm{k}l} \rangle,    \label{OP_l_inter}
\end{align} 
where $\bar{l}$ indicates the layer different from $l$, i.e., $\{l,\bar{l}\,\} = \{A,B\}$.
The dPI order parameter $\Delta_{l}$ is given by an intralayer term $\Delta_{1 l}$ and an interlayer term $\Delta_{2 \bar{l}}$.
Taking into account the dPI order parameter, we obtain the mean field Hamiltonian 
\begin{equation}
  H^{\rm MF} = \sum_{\bm{k}} \hat{C}^{\dagger}_{\bm{k}} \hat{H}^{\rm MF}_{4} (\bm{k}) \hat{C}_{\bm{k}} + E_{\rm cond}, \label{H_MF}
\end{equation}
where 
\begin{equation}
  E_{\rm cond} = \frac{N}{2 g_{1}} [(\Delta_{1 A})^{2} + (\Delta_{1 B})^{2}] + \frac{N}{g_{2}} \Delta_{2 A} \Delta_{2 B}, \label{E_cond}
\end{equation}
and $\hat{C}^{\dagger}_{\bm{k}} = \left(c^{\dagger}_{\bm{k} \uparrow A}, c^{\dagger}_{\bm{k} \downarrow A}, c^{\dagger}_{\bm{k} \uparrow B}, c^{\dagger}_{\bm{k} \downarrow B}\right)$.
The $4 \times 4$ matrix $\hat{H}^{\rm MF}_{4} (\bm{k})$ is given by
\begin{equation}
  \hat{H}^{\rm{MF}}_{4} (\bm{k}) =
  \scalebox{1.2}{$\displaystyle
    {\footnotesize
      \begin{pmatrix}
        \xi_{\bm{k} A}                & -\alpha \lambda_{\bm{k}}^{+}      & t_{\perp}                         & 0 \\
        -\alpha \lambda_{\bm{k}}^{-}  & \xi_{\bm{k} A}                    & 0                               & t_{\perp} \\
        t_{\perp}                    & 0                              & \xi_{\bm{k} B}                     &  \alpha \lambda_{\bm{k}}^{+} \\
        0                          & t_{\perp}                        &  \alpha \lambda_{\bm{k}}^{-}       & \xi_{\bm{k} B}  
      \end{pmatrix} ,
    } $}
  \label{H_MF_matrix}
\end{equation}
with $\lambda_{\bm{k}}^{\pm} = \sin k_{y} \pm i \sin k_{x}$ 
and $\xi_{\bm{k} l} = \varepsilon_{\bm{k}} + d_{\bm{k}} \Delta_{l}$.
The matrix $\hat{H}^{\rm MF}_{4} (\bm{k})$ is diagonalized by a unitary transformation, 
and we obtain the mean field Hamiltonian in the band representation
\begin{equation}
  H^{\rm MF} = \sum_{\bm{k}} \sum^{4}_{a = 1} E_{\bm{k} a} \gamma^{\dagger}_{\bm{k} a} \gamma_{\bm{k} a} + E_{\rm cond}, \label{H_MF_BAND}
\end{equation}
where $E_{\bm{k} a}$ is an energy dispersion of the $a$-th eigenstate.
Then, Eqs. (\ref{OP_l_intra}) and (\ref{OP_l_inter}) are reduced to
\begin{align}
  \Delta_{1 l} &= - \frac{g_{1}}{N} \sum_{\bm{k}, s} \sum^{4}_{a = 1} d_{\bm{k}} |u^{a}_{\bm{k} s l}|^{2} f(E_{\bm{k} a}),    \label{OP_l_intra_BAND} \\
  \Delta_{2 l} &= - \frac{g_{2}}{N} \sum_{\bm{k}, s} \sum^{4}_{a = 1} d_{\bm{k}} |u^{a}_{\bm{k} s l}|^{2} f(E_{\bm{k} a}),    \label{OP_l_inter_BAND}
\end{align}
where $f(E)$ is the Fermi-Dirac distribution function,
and $u^{a}_{\bm{k} s l}$ are elements of the unitary matrix defined by 
\begin{equation}
  c_{\bm{k} s l} = \sum_{a=1}^{4} u^{a}_{\bm{k} s l} \gamma_{\bm{k} a}.   \label{Unitary_Transformation}
\end{equation}
We numerically solve self-consistent equations for the dPI order parameters, namely, Eqs. (\ref{OP_l_intra_BAND}) and (\ref{OP_l_inter_BAND}).
The thermodynamically stable state is determined by calculating free energies of the normal state, EQ state, and EO state.

Let us show explicit forms of the unitary matrix and quasiparticle's energy.
The matrix $\hat{U} (\bm{k})$ is given by
\begin{align}
  \hat{U} & (\bm{k}) = \frac{1}{\sqrt{2}} \times \notag \\
  &
  \scalebox{0.9}{$\displaystyle
    {\footnotesize
      \begin{pmatrix}
        T_{\bm{k -}}  &  -\frac{\lambda_{\bm{k}}^{+}}{|\bm{g}_{\bm{k}}|} \sqrt{1 - T_{\bm{k +}}^{2}}  &  \sqrt{1 - T_{\bm{k} -}^{2}}  &  -\frac{\lambda_{\bm{k}}^{+}}{|\bm{g}_{\bm{k}}|} T_{\bm{k} +} \\
        \\
        \frac{\lambda_{\bm{k}}^{-}}{|\bm{g}_{\bm{k}}|} T_{\bm{k} -}  &  \sqrt{1 - T_{\bm{k} +}^{2}}  &  \frac{\lambda_{\bm{k}}^{-}}{|\bm{g}_{\bm{k}}|} \sqrt{1 - T_{\bm{k} -}^{2}}  &  T_{\bm{k} +} \\
        \\
        \sqrt{1 - T_{\bm{k} -}^{2}}  &  -\frac{\lambda_{\bm{k}}^{+}}{|\bm{g}_{\bm{k}}|} T_{\bm{k} +}  &  -T_{\bm{k} -}  &  \frac{\lambda_{\bm{k}}^{+}}{|\bm{g}_{\bm{k}}|}  \sqrt{1 - T_{\bm{k} +}^{2}} \\
        \\
        \frac{\lambda_{\bm{k}}^{-}}{|\bm{g}_{\bm{k}}|}  \sqrt{1 - T_{\bm{k} -}^{2}}  &  T_{\bm{k} +}  &  -\frac{\lambda_{\bm{k}}^{-}}{|\bm{g}_{\bm{k}}|} T_{\bm{k} -}  &  -\sqrt{1 - T_{\bm{k} +}^{2}}
      \end{pmatrix} ,
    } $}
  \label{unitary_matrix}
\end{align}
where $|\bm{g}_{\bm k}| = (\sin^{2} k_{x} + \sin^{2} k_{y})^{1/2}$ is the magnitude of the Rashba g-vector,
and $T_{\bm{k} \pm}$ is defined by
\begin{equation}
  T_{\bm{k} \pm} = \frac{t_{\perp}}{\sqrt{t^{2}_{\perp} + \left[\alpha^{d}_{\bm{k} \pm} + \sqrt{ \left(\alpha^{d}_{\bm{k} \pm}\right)^{2} + t^{2}_{\perp}}\right]^{2}}}. \label{T_pm}
\end{equation}
We have introduced $\alpha^{d}_{\bm{k} \pm}$ as
\begin{equation}
  \alpha^{d}_{\bm{k} \pm} = \alpha |\bm{g}_{\bm{k}}| \pm d_{\bm{k}} \frac{\Delta_{A} - \Delta_{B}}{2}. \label{ASOC_dPI}
\end{equation}
The energy dispersion is obtained as
\begin{align}
  E_{\bm{k} 1} &= \varepsilon_{\bm{k}} + d_{\bm{k}} \frac{\Delta_{A} + \Delta_{B}}{2} + \sqrt{(\alpha^{d}_{\bm{k} -})^{2} + t^{2}_{\perp}} , \label{E1_BAND} \\
  E_{\bm{k} 2} &= \varepsilon_{\bm{k}} + d_{\bm{k}} \frac{\Delta_{A} + \Delta_{B}}{2} + \sqrt{(\alpha^{d}_{\bm{k} +})^{2} + t^{2}_{\perp}} , \label{E2_BAND} \\
  E_{\bm{k} 3} &= \varepsilon_{\bm{k}} + d_{\bm{k}} \frac{\Delta_{A} + \Delta_{B}}{2} - \sqrt{(\alpha^{d}_{\bm{k} -})^{2} + t^{2}_{\perp}} , \label{E3_BAND} \\
  E_{\bm{k} 4} &= \varepsilon_{\bm{k}} + d_{\bm{k}} \frac{\Delta_{A} + \Delta_{B}}{2} - \sqrt{(\alpha^{d}_{\bm{k} +})^{2} + t^{2}_{\perp}} . \label{E4_BAND}
\end{align}

\subsection{Electric multipole order}
In this subsection, we identify electric multipole order. For simplicity, a tetragonal system with $\delta t_{1} = 0$ is considered.
Then, the bilayer Rashba system preserves $D_{4h}$ point group symmetry. 
The dPI order parameter on the $l$-layer is Ising-type, namely, $\Delta_{l} = \Delta$ or $-\Delta$.
By taking account of layer degrees of freedom, we can identify two types of electric multipole states 
from the viewpoint of symmetry.

\begin{figure}[htbp]
 \begin{center}
   \includegraphics[width=0.80\hsize]{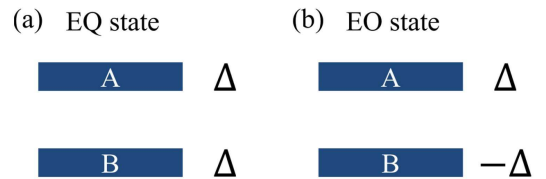}
   \caption{(Color online) Illustration of electric multipole states in the bilayer system.
     The blue lines indicate bilayers,
     and the dPI order parameter on each layer is shown by $\Delta$ or $-\Delta$.
     (a) The EQ state by ferroic stacking $(\Delta_{A}, \Delta_{B}) = (\Delta, \Delta)$.
     (b) The EO state due to antiferroic stacking $(\Delta_{A}, \Delta_{B}) = (\Delta, -\Delta)$.
   }
   \label{multipole_state}
 \end{center}
\end{figure}

When the dPI order parameter is ferroically stacked between bilayers, i.e., $(\Delta_{A}, \Delta_{B}) = (\Delta, \Delta)$,
the state is identified as an even-parity EQ state [Fig.~\ref{multipole_state}(a)].
The EQ state is characterized by a finite EQ moment with $O_{x^{2} - y^{2}}$ symmetry, 
and symmetry of the system is reduced to $D_{2h}$ point group.
Because inversion symmetry and time-reversal symmetry are preserved in the EQ state,
the band structure has spin degeneracy in accordance with Kramers theorem. 
Indeed, the energy of quasiparticles is obtained from Eqs. (\ref{E1_BAND})-(\ref{E4_BAND}) as
\begin{align}
  E_{\bm{k} 1} &= E_{\bm{k} 2} = \varepsilon_{\bm{k}} + d_{\bm{k}} \Delta + \sqrt{(\alpha |\bm{g}_{\bm{k}}|)^{2} + t^{2}_{\perp}}, \label{E1_E2_EQ} \\
  E_{\bm{k} 3} &= E_{\bm{k} 4} = \varepsilon_{\bm{k}} + d_{\bm{k}} \Delta - \sqrt{(\alpha |\bm{g}_{\bm{k}}|)^{2} + t^{2}_{\perp}}. \label{E3_E4_EQ}
\end{align}
The EQ state can also be regarded as the electronic nematic state with broken rotation symmetry~\cite{Yamase_PI_2,Yamase_PI_3}.

On the other hand, an odd-parity EO state is characterized
by antiferroic stacking of the dPI order parameter,
i.e., $(\Delta_{A}, \Delta_{B}) = (\Delta, -\Delta)$ [Fig.~\ref{multipole_state}(b)].
The EO state hosts an EO moment with $T_{(x^{2} - y^{2})z}$ symmetry,
and symmetry of the system is reduced to noncentrosymmetric $D_{2d}$ point group.
Then, global inversion symmetry is spontaneously broken by odd-parity EO order.
The combined effect of broken inversion symmetry and layer-dependent Rashba ASOC
lifts spin degeneracy in the band structure.
From Eqs. (\ref{E1_BAND})-(\ref{E4_BAND}), the energy dispersion is represented as
\begin{align}
  E_{\bm{k} 1} &= \varepsilon_{\bm{k}} + \sqrt{(\alpha |\bm{g}_{\bm{k}}| - d_{\bm{k}} \Delta)^{2} + t^{2}_{\perp}}, \label{E1_EO} \\
  E_{\bm{k} 2} &= \varepsilon_{\bm{k}} + \sqrt{(\alpha |\bm{g}_{\bm{k}}| + d_{\bm{k}} \Delta)^{2} + t^{2}_{\perp}}, \label{E2_EO} \\
  E_{\bm{k} 3} &= \varepsilon_{\bm{k}} - \sqrt{(\alpha |\bm{g}_{\bm{k}}| - d_{\bm{k}} \Delta)^{2} + t^{2}_{\perp}}, \label{E3_EO} \\
  E_{\bm{k} 4} &= \varepsilon_{\bm{k}} - \sqrt{(\alpha |\bm{g}_{\bm{k}}| + d_{\bm{k}} \Delta)^{2} + t^{2}_{\perp}}. \label{E4_EO}
\end{align}
In contrast to the EQ state, the energy spectrum is fourfold symmetric because rotoinversion symmetry is preserved.

\section{Electric Octupole Order in Orthorhombic System}

In this paper we aim to explore electromagnetic responses in the parity-violating EO state. 
Although previous studies focused on the tetragonal system~\cite{Hitomi_EO_JPSJ2014,Hitomi_EO_JPSJ2016}, 
a candidate material, YBCO, is an orthorhombic compound. 
Therefore, in this section we study characteristic properties of the EO state 
in orthorhombic systems and discuss consistency with a recent experiment for YBCO~\cite{Sato_YBCO}.

Since the crystal structure of YBCO contains quasi-one-dimensional CuO chains, 
crystal symmetry belongs to orthorhombic $D_{2h}$ point group.
$C_{4}$ rotation symmetry is slightly broken, that is, weak nematicity appears in the normal state~\cite{Wu_YBCO}.
For such orthorhombic systems we here assume anisotropy in the  energy dispersion, $\delta t_1 \ne 0$. 
Because symmetry of the normal state is equivalent to the EQ state, EQ order can not be a second order phase transition. 
On the other hand, EO order can be a second order phase transition 
since it is accompanied by spontaneous inversion symmetry breaking. 
When EO order occurs, symmetry is reduced from $D_{2h}$ to $C_{2v}$ lacking inversion symmetry. 
In contrast to the tetragonal EO state, the orthorhombic EO state is polar, in the sense that the electric polarization is allowed. Thus, the EO moment coexists with the electric dipole moment.
A set of parameters $(t_{1}, t_{2}, t_{\perp}, \delta t_{1}, \alpha, \mu) = (1.0, -0.25, 0.1, 0.05, 0.05, -0.6)$ 
is consistent  with Fermi surfaces in YBCO.~\cite{Shen_review,Harrison_SOC} 
Thus, the parameter set is adopted in this section, unless mentioned otherwise.  
To discuss the EO state, we set $(g_{1}, g_{2}) = (0.9, -0.25)$.

First, the order parameter and the nematicity are shown in Sec.~3.1. 
Next, the magnetic torque is calculated and compared with experimental results in Sec~3.2.
In Sec.~3.3 we show that anisotropy in the spin susceptibility is significantly enhanced in the superconducting EO state.

\subsection{Order parameter}
\begin{figure}[htbp]
  \begin{center}
\hspace*{-10mm}
    \includegraphics[width=0.80\hsize]{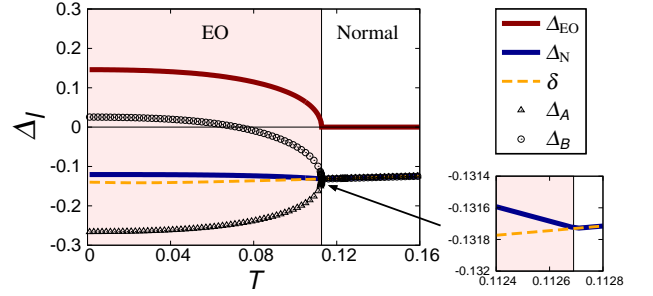}
    \caption{(Color online) The order parameter of the EO state $\Delta_{\rm EO} = (\Delta_A - \Delta_B) / 2$ 
and the nematicity $\Delta_{\rm N} = (\Delta_A + \Delta_B) / 2$ (thick solid lines).
      $\Delta_{A}$ and $\Delta_{B}$ are also shown by triangles and circles, respectively.
      The nematicity in the normal state is depicted by $\delta$ (thin dashed line).
      The EO phase is highlighted by the red color.
      The second order phase transition occurs at around $T_{\rm c} \simeq 0.1127$.
      The lower right panel shows $\Delta_{\rm{N}}$ and $\delta$ around the critical temperature.
    }
    \label{op_YBCO}
  \end{center}
\end{figure}

Let us show the order parameter and the nematicity in Fig.~\ref{op_YBCO}.
The dPI parameter $\Delta_{l}$ defined by Eq.~(\ref{OP_l}) represents the nematicity on the $l$-layer.
Because fourfold rotation symmetry is originally broken in the orthorhombic system, 
both $\Delta_{A}$ and $\Delta_B$ are finite. Thus, $\Delta_{l}$ can not be an order parameter 
of symmetry breaking second order phase transition. 
However, the order parameter of the EO state is well-defined by 
$\Delta_{\rm{EO}} = ( \Delta_{B} - \Delta_{A} ) / 2$, which implies 
the degree of inversion symmetry breaking.
Figure~\ref{op_YBCO} shows appearance of  $\Delta_{\rm{EO}}$ below the critical temperature  $T_{\rm c} \simeq 0.1127$ 
with a mean field critical exponent. Recent experimental studies actually reported the second order phase transition 
at the onset temperature of pseudogap in YBCO~\cite{Sato_YBCO, Shekhter2013}.

We here define the nematicity of the system by $\Delta_{\rm{N}} = (\Delta_A + \Delta_B) / 2$. 
As we see in the lower right panel of  Fig.~\ref{op_YBCO}, the nematicity shows a kink at the critical temperature. 
This is consistent with the fact that the nematicity is not a primary order parameter, 
but it is coupled with the EO order.

\subsection{Magnetic torque and pseudogap}
A change in the nematicity accompanied with the second order EO phase transition may be consistent with 
the recent magnetic torque measurement~\cite{Sato_YBCO}. 
In order to examine the consistency we calculate the spin susceptibility given by
\begin{equation}
  \chi_{\mu \nu} (\bm{q} , i\omega_{n}) = \frac{1}{N} \int^{1/T}_{0} d \tau \hspace{0.5mm} \langle T_{\tau} \{ S_{\mu} (- \bm{q} ,\tau) \hspace{0.3mm} S_{\nu} ( \bm{q} , 0) \} \rangle \hspace{0.5mm} e^{i \omega_{n} \tau} , \label{spin_susceptibility}
\end{equation}
where $S_{\mu} (\bm{q}) = \sum_{\bm{k}} \sum_{s,s',l,l'} \langle s l | \hat{S}_{\mu} | s' l' \rangle c^{\dagger}_{\bm{k} + \bm{q} s l} c_{\bm{k} s' l'}$ for $\mu=x,y$, and $\omega_{n} = 2n \pi T$ are boson Matsubara frequencies.
We now consider the uniform and static spin susceptibility and take the limit $(\bm{q}, \omega_{n}) \rightarrow (\bm{0}, 0)$.
Then, the spin susceptibility is obtained as
\begin{equation}
  \chi_{\mu \nu} = \chi^{\rm P}_{\mu \nu} + \chi^{\rm VV}_{\mu \nu} , \label{spin_susceptibility_Pauli_and_Van_Vleck}
\end{equation}
where
\begin{align}
  \chi^{\rm P}_{\mu \nu}  &= \frac{1}{N} \sum_{E_{\bm{k} a} = E_{\bm{k} b}} \sum_{\bm{k}} [\tilde{S}_{\mu} (\bm{k})]_{ab} [\tilde{S}_{\nu} (\bm{k})]_{ba} \{-f'(E_{\bm{k} a}) \}, \label{spin_susceptibility_Pauli} \\
  \chi^{\rm VV}_{\mu \nu} &= -\frac{1}{N} \sum_{E_{\bm{k} a} \neq E_{\bm{k} b}} \sum_{\bm{k}} [\tilde{S}_{\mu} (\bm{k})]_{ab} [\tilde{S}_{\nu} (\bm{k})]_{ba} \frac{f(E_{\bm{k} a})-f(E_{\bm{k} b})}{E_{\bm{k} a} - E_{\bm{k} b}}, \label{spin_susceptibility_Van_Vleck}
\end{align}
with $[\tilde{S}_{\mu} (\bm{k})]_{ab} = \langle a | S_{\mu} (\bm{k}) | b \rangle$ being spin operator in the band representation. 
We used the derivative of the Fermi-Dirac distribution function $f' (E)$.
The spin susceptibility $\chi_{\mu \nu}$ is separated into the Pauli part $\chi^{\rm P}_{\mu \nu}$ and the Van Vleck part $\chi^{\rm VV}_{\mu \nu}$.
The Pauli susceptibility originates from the intraband contribution,
while the Van Vleck susceptibility is given by the interband contribution.

By calculating $[\tilde{S}_{\mu} (\bm{k})]_{ab}$,
we obtain the analytic form of the diagonal spin susceptibility $\chi_{\mu \mu}$ as 
\begin{align}
  \chi^{\rm P}_{\mu \mu}  &= \frac{g^{2} \mu^{2}_{\rm B}}{4 N} \sum_{\bm{k},a} \frac{\sin^{2} k_{\bar{\mu}} + \{T^{(2)} (\bm{k})\}^{2} \sin^{2} k_{\mu}}{|\bm{g}_{\bm{k}}|^{2}} \{ - f'(E_{\bm{k a}}) \}, \label{spin_susceptibility_diagonal_Pauli_EQ} \\
  \chi^{\rm VV}_{\mu \mu} &= -\frac{g^{2} \mu^{2}_{\rm B}}{2 N} \sum_{\bm{k}} \frac{\{T^{(1)} (\bm{k})\}^{2} \sin^{2} k_{\mu}}{|\bm{g}_{\bm{k}}|^{2}} \hspace{1mm} \{ F_{14} (\bm{k}) + F_{23} (\bm{k}) \}, \label{spin_susceptibility_diagonal_Van_Vleck_EQ}
\end{align} 
in the normal state, 
while $\chi_{\mu \mu}$ in the EO state is obtained as
\begin{align}
  \chi^{\rm P}_{\mu \mu}  &= \frac{g^{2} \mu^{2}_{\rm B}}{4 N} \sum_{\bm{k},a} \frac{\sin^{2} k_{\bar{\mu}}}{|\bm{g}_{\bm{k}}|^{2}} \hspace{1mm} \{ - f'(E_{\bm{k a}}) \} , \label{spin_susceptibility_diagonal_Pauli_EO} \\
  \chi^{\rm VV}_{\mu \mu} &= -\frac{g^{2} \mu^{2}_{\rm B}}{2 N} \sum_{\bm{k}} \frac{\sin^{2} k_{\mu}}{|\bm{g}_{\bm{k}}|^{2}} \biggl[ \{T^{(1)} (\bm{k})\}^{2} \{ F_{14} (\bm{k}) + F_{23} (\bm{k}) \} \notag \\
    & \hspace{26mm} + \{T^{(2)} (\bm{k})\}^{2} \{ F_{12} (\bm{k}) + F_{34} (\bm{k}) \}\biggr]. \label{spin_susceptibility_diagonal_Van_Vleck_EO} 
\end{align}
Here $\bar{\mu}$ indicates the direction orthogonal to $\mu$, i.e., $\{ \mu, \bar{\mu} \} = \{ x, y \}$.
We have introduced $T^{(1)} (\bm{k})$, $T^{2}(\bm{k})$, and $F_{ab} (\bm{k})$ by
\begin{align}
  T^{(1)} (\bm{k}) &= T_{\bm{k} -} T_{\bm{k} +} - \sqrt{1 - T^{2}_{\bm{k} -}} \sqrt{1 - T^{2}_{\bm{k} +}}, \label{T1_k} \\
  T^{(2)} (\bm{k}) &= T_{\bm{k} -} \sqrt{1 - T^{2}_{\bm{k} +}} + T_{\bm{k} +} \sqrt{1 - T^{2}_{\bm{k} -}}, \label{T2_k} \\
  F_{ab} (\bm{k}) &= \frac{f(E_{\bm{k} a}) - f(E_{\bm{k} b})}{E_{\bm{k} a} - E_{\bm{k} b}}. \label{F_ab}
\end{align}

While Kramers theorem ensures $E_{\bm{k} 1} = E_{\bm{k} 2}$ and $E_{\bm{k} 3} = E_{\bm{k} 4}$ in the normal state, 
such twofold degeneracy in the electronic band structure is lifted in the EO state.  
The second term of the Pauli susceptibility in Eq.~(\ref{spin_susceptibility_diagonal_Pauli_EQ}) comes from the two-fold degeneracy. 
This term indeed disappears in the EO state [Eq.~(\ref{spin_susceptibility_diagonal_Pauli_EO})].
Hence, a part of the Pauli susceptibility in the normal state changes to the Van Vleck susceptibility in the EO state,
as shown by the second term in Eq.~(\ref{spin_susceptibility_diagonal_Van_Vleck_EO}).
This is important for the spin susceptibility in the superconducting state (see Sec.~3.3).

\begin{figure}[htbp]
 \begin{center}
   \includegraphics[width=0.85\hsize]{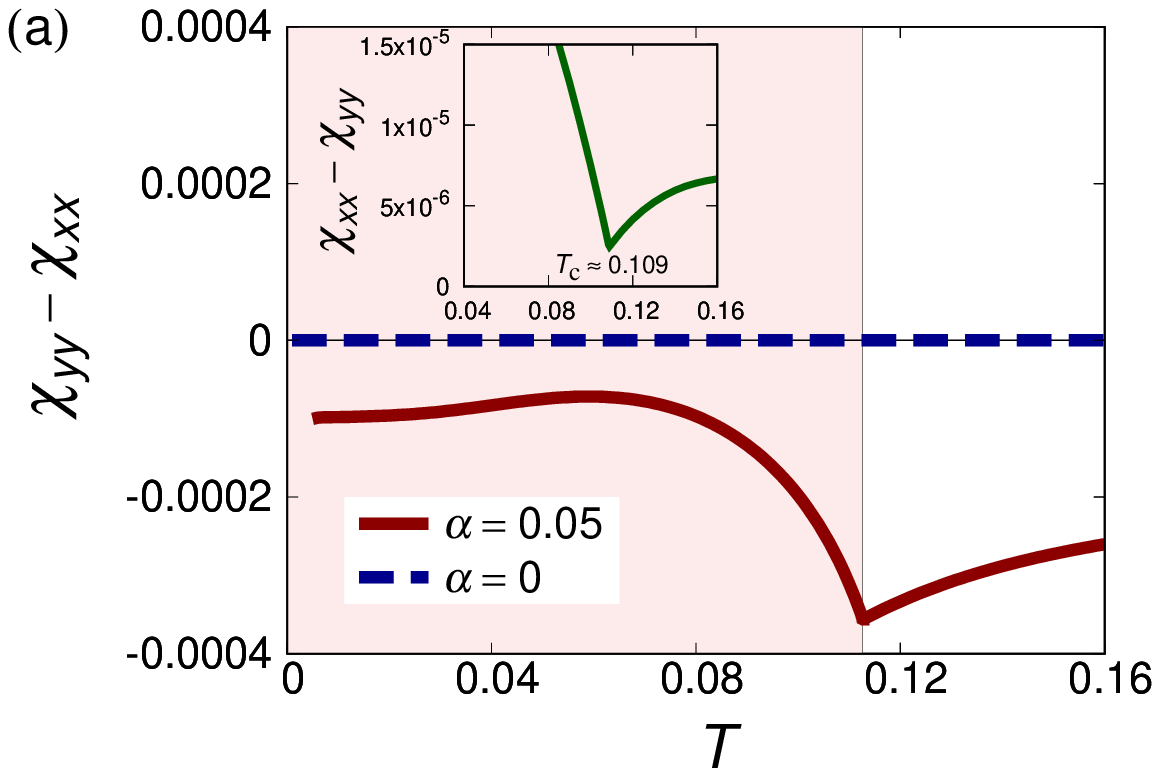}
\\
\vspace{5mm}
   \includegraphics[width=0.75\hsize]{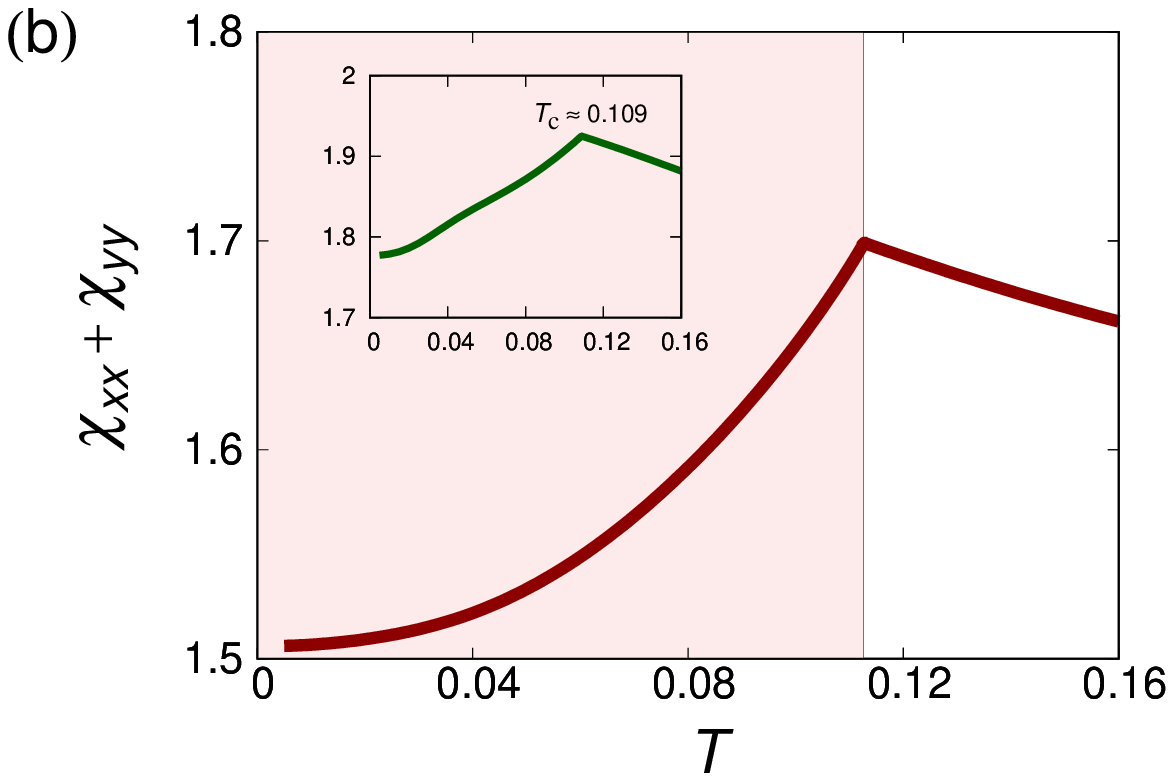}
   \caption{(Color online) (a) Temperature dependence of the magnetic torque, i.e., $\chi_{yy} - \chi_{xx}$.
     The magnetic torque shows a kink for $\alpha = 0.05$ (red solid line), although it vanishes 
     at $\alpha = 0$ (blue dashed line).
     In both cases, the second order phase transition of EO order occurs at around $T_{\rm c} \simeq 0.1127$.
     (b) In-plane spin susceptibility $\chi_{xx} + \chi_{yy}$ as a function of the temperature.
     In both panels, inset shows the results for $(\delta t_{1}, \alpha, \mu, g_{1}) = (0.02, 0.01, -0.75, 0.7)$ while the other parameters kept.
   }
   \label{torque_and_pseudogap}
 \end{center}
\end{figure}

Figure~\ref{torque_and_pseudogap}(a) shows the calculated temperature dependence of the magnetic torque, i.e., $\chi_{yy} - \chi_{xx}$.
Owing to the layer-dependent Rashba ASOC term, the spin susceptibility is anisotropic in the whole temperature region.
We find a pronounced kink at around $T_{\rm{c}} \simeq 0.1127$ consistent with experimental observation~\cite{Sato_YBCO}. 
The kink appears at the transition temperature of EO order. 
A similar change in the nematicity was also observed in the Nernst effect~\cite{Daou2010}. 
The magnetic torque shows linear temperature dependence near the critical point, 
$\chi_{yy}(T) - \chi_{xx}(T) - \left[ \chi_{yy}(T_{\rm c}) - \chi_{xx}(T_{\rm c}) \right] \propto T_{\rm{c}} - T$, 
because the nematicity is a secondary order parameter of EO order. On the other hand, when EQ order occurs in a tetragonal system, 
square root behavior $\propto (T_{\rm c}-T)^{1/2}$ is expected below $T_{\rm c}$, which disagrees with the experiment. 

Although the kink at the critical point of EO order is a generally obtained, detailed temperature dependence 
of the magnetic torque depends on parameters of the model. For small $\delta t_1=0.02$ and $\alpha=0.01$, 
the magnitude of the magnetic torque increases below the EO transition temperature [inset of Fig.~\ref{torque_and_pseudogap}(a)]. 
This temperature dependence is consistent with the magnetic torque measurement~\cite{Sato_YBCO}. 
Consistency with out-of-plane magnetic anisotropy is also shown in Appendix B.  
Note that the magnetic torque disappears in the absence of spin-orbit coupling. 
At $\alpha=0$, indeed, we obtain an isotropic form 
\begin{equation}
 \chi_{\mu \mu} = \frac{g^{2} \mu^{2}_{\rm B}}{4 N} \sum_{\bm{k}} \sum^{4}_{a = 1} \hspace{1mm} \{- f' (E_{\bm{k} a})\}, \label{spin_susceptibility_diagonal_Pauli_ASOC_zero_result}
\end{equation}
which leads to $\chi_{yy} - \chi_{xx} =0$.
The layer-dependent Rashba spin-orbit coupling plays an essential role for the kink in the magnetic torque.

We also show temperature dependence of the spin susceptibility $\chi_{xx} + \chi_{yy}$ 
in Fig.~\ref{torque_and_pseudogap}(b).
The spin susceptibility shows a peak at the transition temperature and 
decreases with temperature below $T_{\rm c}$, consistent with the pseudogap behavior 
observed in YBCO~\cite{Walstedt_YBCO,Warren_YBCO}. 
This pseudogap phenomenon occurs because the density of states is decreased in the EO state.

\subsection{Enhanced anisotropy in superconducting state}

\begin{figure}[htbp]
 \begin{center}
   \includegraphics[width=0.80\hsize]{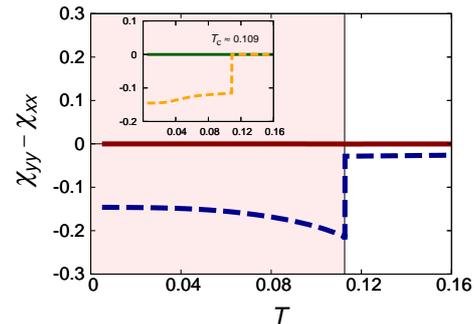}
   \caption{(Color online) Temperature dependence of anisotropy in the Van Vleck spin susceptibility 
$\chi^{\rm VV}_{yy} - \chi^{\rm VV}_{xx}$ (dashed line). Total spin susceptibility $\chi_{yy} - \chi_{xx}$ is again shown 
for a comparison (solid line). 
 Inset shows the results for $(\delta t_{1}, \alpha, \mu, g_{1}) = (0.02, 0.01, -0.75, 0.7)$.
   }
   \label{torque_total_and_VV}
 \end{center}
\end{figure}

Here we discuss the effect of EO order in the superconducting state. 
For this purpose we show the Van Vleck spin susceptibility. 
Figure~\ref{torque_total_and_VV} shows in-plane anisotropy of the Van Vleck spin susceptibility 
$\chi^{\rm VV}_{yy} - \chi^{\rm VV}_{xx}$. Comparison with $\chi_{yy} - \chi_{xx}$ plotted in the same figure 
reveals a remarkably large in-plane anisotropy in the Van Vleck spin susceptibility, that is furthermore enhanced 
in the EO state.

The Van Vleck spin susceptibility coincides with the spin susceptibility in the spin-singlet superconducting state at $T=0$ 
when the band splitting energy is much lager than the superconducting gap.~\cite{Yanase2007b} 
Therefore, the result in Fig.~\ref{torque_total_and_VV} indicates that in-plane anisotropy of spin susceptibility 
is significantly enhanced in the superconducting state. When the superconductivity occurs in the EO state,  
the enhancement factor reaches $1000$ almost irrespective of choices of parameters. 
For small $\delta t_1=0.02$ and $\alpha=0.01$ adopted in the inset of Fig.~\ref{torque_total_and_VV}, 
$\chi^{\rm VV}_{yy} - \chi^{\rm VV}_{xx}$ is more than 5000 times  larger than $\chi_{yy} - \chi_{xx}$.

Another care is required for discussion of high-temperature cuprate superconductors because the 
superconducting gap may be larger than the spin splitting energy $\Delta_{\rm so}$ due to EO order. 
Then, the Van Vleck spin susceptibility decreases in the superconducting state. 
Assuming $\Delta =30$meV, $\alpha=5$meV,~\cite{Gotlieb2018,Harrison_SOC} $t_\perp =25$meV, 
and the superconducting gap $\Delta_{\rm sc}=30$meV,~\cite{Hashimoto_review} 
we have the ratio $\Delta_{\rm so}/\Delta_{\rm sc} \sim 1/3$. 
Although cuprates are $d$-wave superconductors, 
we use the result for $s$-wave superconductors~\cite{Frigeri_chi} for a rough estimation.
Then, we estimate that in-plane anisotropy of spin susceptibility $\chi_{yy} - \chi_{xx}$ in the superconducting EO state is 100 times 
larger than that in the normal state for $\delta t_1=0.02$ and $\alpha=0.01$. 
Such a large anisotropy may be detected by a nuclear magnetic resonance (NMR) experiment. 

Furthermore, experimental studies of anisotropic spin susceptibility in the superconducting state may detect the EO order. 
For the above parameters, anisotropy in the superconducting state is five times enhanced by EO order. 
Therefore, the anisotropy decreases with increasing hole doping in the superconducting state 
if EO order disappears at the quantum critical point. 

\begin{figure}[htbp]
 \begin{center}
\hspace{-10mm}
   \includegraphics[width=0.70\hsize]{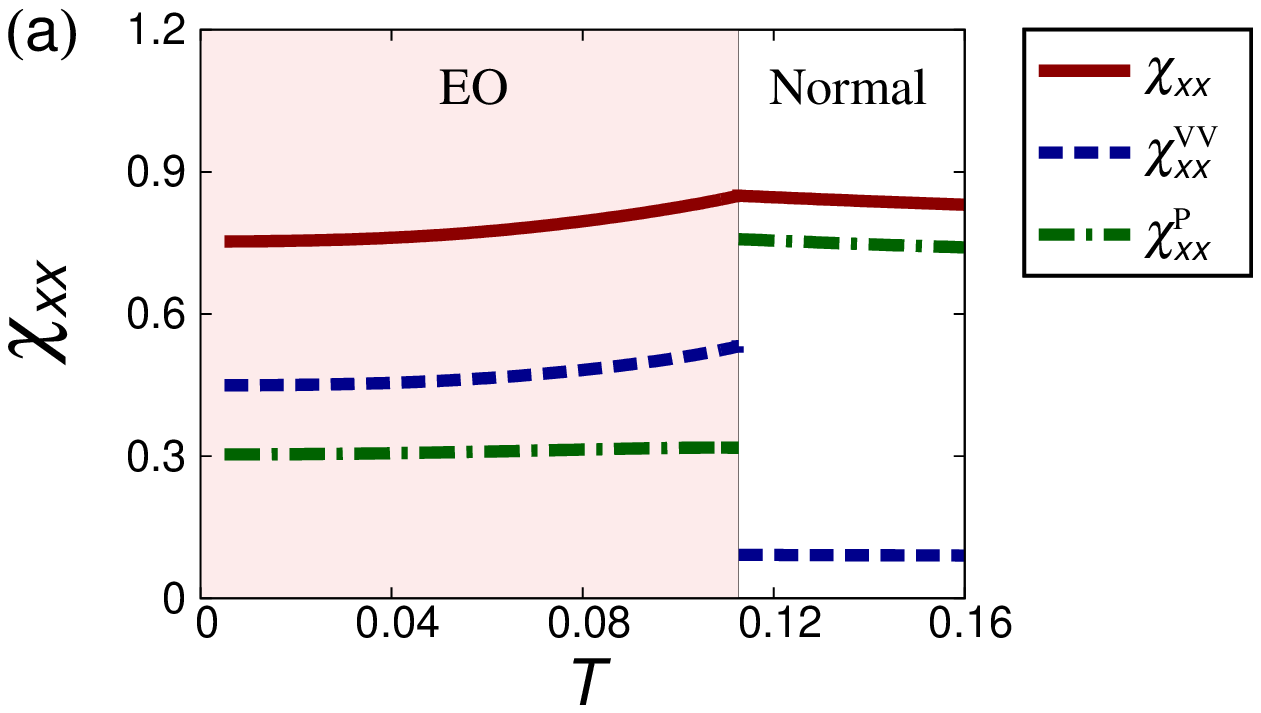}
\\
\vspace{5mm}
\hspace{-10mm}
   \includegraphics[width=0.70\hsize]{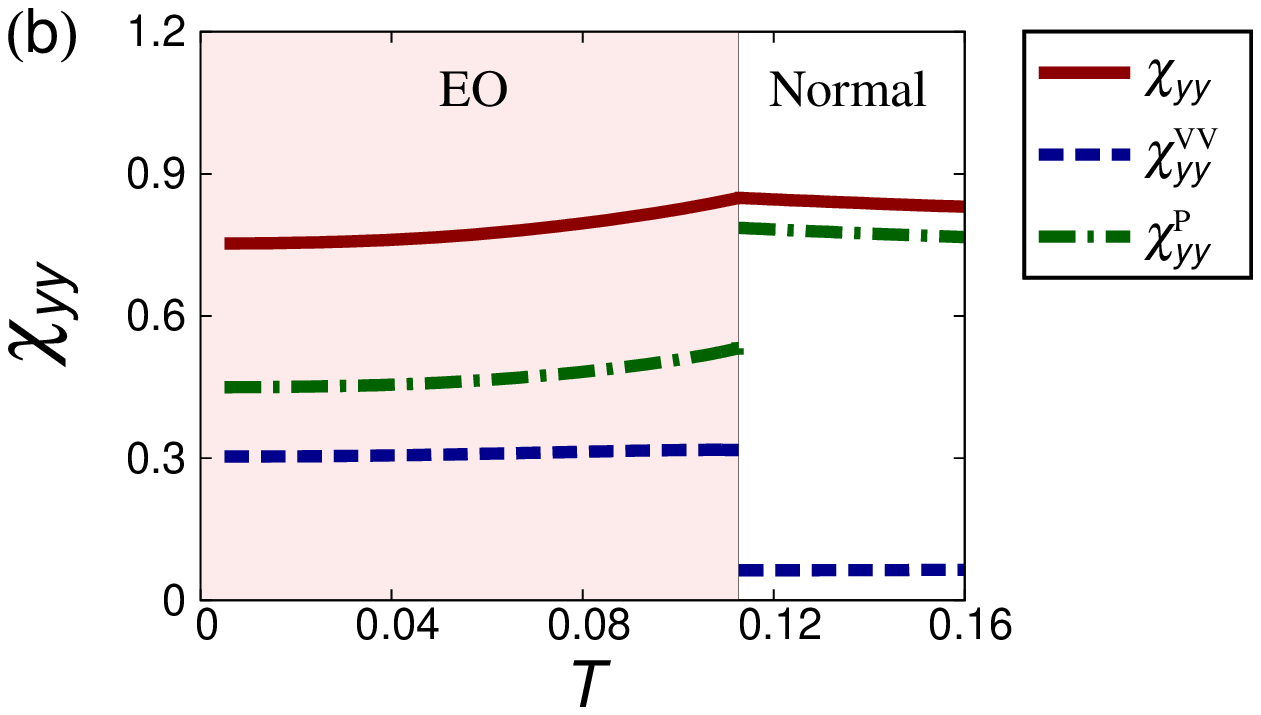}
   \caption{(Color online) Temperature dependence of the spin susceptibility, (a) $\chi_{xx}$ and (b) $\chi_{yy}$.
     The red solid, blue dashed, and green dash-dotted lines show $\chi_{\mu \mu}$, $\chi^{\rm VV}_{\mu \mu}$, 
and $\chi^{\rm P}_{\mu \mu}$, respectively.
   }
   \label{spin_susceptibilities}
 \end{center}
\end{figure}

We show $\chi_{\mu \mu}$, $\chi^{\rm VV}_{\mu \mu}$, and $\chi^{\rm P}_{\mu \mu}$ for $\mu =x, y$ 
in Fig.~\ref{spin_susceptibilities}. 
Although $\chi_{xx}$ is nearly the same as $\chi_{yy}$, contributions from Van Vleck and Pauli parts are 
significantly different between $\chi_{xx}^{\rm VV/P}$ and $\chi_{yy}^{\rm VV/P}$. 
The Van Vleck part is increased by EO order because the intraband contribution changes to 
the interband contribution as a result of the spin splitting in the band structure. 
Therefore, the anisotropy in the Van Vleck spin susceptibility is also enhanced by EO order.

Finally, we show the symmetry of superconductivity in the EO state. In the orthorhombic crystal, 
the $d_{x^2-y^2}$-wave superconductivity belongs to the $A_g$ representation of $D_{2h}$. 
This means that the $d$-wave Cooper pairs coexist with the $s$-wave Cooper pairs. 
In the EO state, these even-parity Cooper pairs mix to the odd-parity ones owing to the inversion symmetry breaking. 
From the compatibility relation, we find that the $p$-wave Cooper pairs in the $B_{1u}$ representation, $k_y \hat{x}$ and $k_x \hat{y}$, 
are mixed. Thus, all these $s$-wave, $p$-wave, and $d$-wave Cooper pairs appear in the superconducting EO state.  
However, the mixing of order parameter does not significantly affect the magnetic properties discussed above.~\cite{Yanase2007b}

\section{Spin Hall Effect}

Hereafter, we show magnetoelectric transport phenomena in the EO state with spontaneous parity violation.
The spin Hall effect is investigated in this section, while the Edelstein effect is studied in the next section. 
We consider the tetragonal and orthorhombic systems.

Let us introduce the spin Hall conductivity.
In the linear-response region, the spin Hall effect~\cite{Murakami_SHE,Sinova_SHE,Inoue_SHE,Kontani_SHE,Mizoguchi_SHE} is described as
\begin{equation}
  J^{s z}_{\mu} = \sigma^{\rm{SHE}}_{\mu \nu} \hspace{0.5mm} E_{\nu}, \label{spin_Hall_effect}
\end{equation}
where $J^{s z}_{\mu}$ is the spin current along the $\mu$-direction with magnetization in the $z$-direction,
$\sigma^{\rm{SHE}}_{\mu \nu}$ is the spin Hall conductivity,
and $E_{\nu}$ is the electric field along the $\nu$-direction.
We here calculate the spin Hall conductivity on the basis of the mean field Hamiltonian [Eq.~(\ref{H_MF})], that is given by the Kubo formula
\begin{equation}
  \sigma^{\rm{SHE}}_{\mu \nu} = \lim_{\omega \rightarrow 0} \frac{K^{\rm{SHE}}_{\mu \nu} (\omega) - K^{\rm{SHE}}_{\mu \nu} (0)}{i \omega} , \label{SHC_kubo}
\end{equation}
\begin{equation}
  K^{\rm{SHE}}_{\mu \nu} (\omega) = \left. K^{\rm{SHE}}_{\mu \nu} (i \omega_{n}) \right|_{i \omega_{n} \rightarrow \omega + i \hspace{0.5mm} 0} . \label{SHC_response_function_kaiseki}
\end{equation}
The correlation function is given by
\begin{equation}
  K^{\rm{SHE}}_{\mu \nu} (i \omega_{n}) = \frac{1}{N} \int^{1/T}_{0} d \tau \hspace{0.5mm} \langle T_{\tau} \{ J^{s z}_{\mu} (\tau) \hspace{0.5mm} J_{\nu} (0) \} \rangle \hspace{0.5mm} e^{i \omega_{n} \tau} , \label{SHC_response_function}
\end{equation}
where $J_{\nu}$ is the charge current operator
\begin{equation}
  J_{\nu} = e \sum_{\bm{k}} \hat{C}^{\dagger}_{\bm{k}} \hat{v}_{\bm{k} \nu} \hat{C}_{\bm{k}} , \label{charge_current}
\end{equation}
with  $\hat{v}_{\bm{k} \nu} = \partial \hat{H}^{\rm MF}_{4} (\bm{k}) / \partial k_{\nu}$. 
The spin current operator is defined as 
\begin{equation}
  J^{s z}_{\mu} = \frac{1}{2} g \mu_{\rm{B}} \sum_{\bm{k}} \hat{C}^{\dagger}_{\bm{k}} \frac{\left\{ \hat{v}_{\bm{k} \mu} , \hat{\sigma}^{\,z}_{4} \right\}}{2}  \hat{C}_{\bm{k}} , \label{spin_current}
\end{equation}
with $g$-factor $g$, Bohr magneton $\mu_{\rm{B}}$, and 
\begin{equation}
  \hat{\sigma}^{\,z}_{4}  =  \scalebox{1.3}{$\displaystyle
    {\footnotesize
      \begin{pmatrix}
        \sigma^{\,z} & 0  \\
        0         & \sigma^{\,z}
      \end{pmatrix}
    } $}. \label{Pauli_matrix_z} 
\end{equation}

Using the mean field Hamiltonian in the band basis [Eq.~(\ref{H_MF_BAND})], 
we obtain the analytic form of the spin Hall conductivity, which is separated into the two terms, 
\begin{equation}
  \sigma^{\rm{SHE}}_{\mu \nu} = \sigma^{{\rm SHE}(\rm C)}_{\mu \nu} + \sigma^{{\rm SHE}(\rm NC)}_{\mu \nu} . \label{SHC_C_and_NC}
\end{equation}
The spin Hall conductivity comes from the interband contributions to the correlation function. 
The first term $\sigma^{{\rm SHE}(\rm C)}_{\mu \nu}$ originates from band splitting between 
$\{E_{\bm{k}1}, E_{\bm{k}2}\}$ and $\{E_{\bm{k}3}, E_{\bm{k}4}\}$. This term is finite unless $\alpha = 0$. 
Thus, the local parity violation gives rise to the spin Hall effect even when the global inversion symmetry is preserved. 
On the other hand, the second term $\sigma^{{\rm SHE}(\rm NC)}_{\mu \nu}$ results from lifting of band degeneracy due to parity violation. 
Since the EO state spontaneously breaks inversion symmetry, spin-orbit coupling gives the second term 
$\sigma^{{\rm SHE}(\rm NC)}_{\mu \nu}$.
The analytic form for $\sigma^{\rm{SHE}}_{xy}$ is 
\begin{align}
  \sigma^{{\rm SHE}(\rm C)}_{xy} &= - \frac{\alpha g \mu_{\rm{B}} e}{N} \sum_{\bm{k}} \frac{\sin k_{x} \cos k_{y}}{|\bm{g}_{\bm{k}}|} \notag \\
  &\hspace{-5mm} \times \Biggl[ \biggl( \frac{\partial \varepsilon_{\bm{k}}}{\partial k_{x}} \biggr) \biggl\{ D_{14} (\bm{k}) + D_{23} (\bm{k}) \biggr\} \biggl\{ T^{2}_{\bm{k} -} + T^{2}_{\bm{k} +} - 1 \biggr\}  \notag \\
    & \hspace{-0mm} - \sin k_{x} \hspace{0.7mm} \biggl\{ D_{14} (\bm{k}) T^{(1)}_{AB} (\bm{k}) + D_{23} (\bm{k}) T^{(1)}_{BA} (\bm{k}) \biggr\}  \notag \\
    & \hspace{-0mm} \times \biggl\{ T_{\bm{k} -} T_{\bm{k} +} + \sqrt{1 - T^{2}_{\bm{k} -}} \sqrt{1 - T^{2}_{\bm{k} +}}  \biggr\} \Biggr] , \label{SHC_xy_C} \\
  \sigma^{{\rm SHE}(\rm NC)}_{xy} &= - \frac{\alpha g \mu_{\rm{B}} e}{N} \sum_{\bm{k}} \frac{\sin k_{x} \cos k_{y}}{|\bm{g}_{\bm{k}}|} \notag \\
  &\hspace{-5mm} \times \Biggl[ \biggl( \frac{\partial \varepsilon_{\bm{k}}}{\partial k_{x}} \biggr) \biggl\{ D_{12} (\bm{k}) - D_{34} (\bm{k}) \biggr\} \biggl\{ T^{2}_{\bm{k} -} - T^{2}_{\bm{k} +} \biggr\}  \notag \\
    & \hspace{-0mm} - \sin k_{x} \hspace{0.7mm} \biggl\{ D_{12} (\bm{k}) T^{(2)}_{AB} (\bm{k}) - D_{34} (\bm{k}) T^{(2)}_{BA} (\bm{k}) \biggr\} \notag \\
    & \hspace{-0mm} \times \biggl\{ T_{\bm{k} -} \sqrt{1 - T^{2}_{\bm{k} +}} - T_{\bm{k} +} \sqrt{1 - T^{2}_{\bm{k} -}} \biggr\} \Biggr] . \label{SHC_xy_NC}
\end{align}
Similarly, the spin Hall conductivity $\sigma^{\rm{SHE}}_{yx}$ is expressed as
\begin{align}
  \sigma^{{\rm SHE}(\rm C)}_{yx} &= \frac{\alpha g \mu_{\rm{B}} e}{N} \sum_{\bm{k}} \frac{\cos k_{x} \sin k_{y}}{|\bm{g}_{\bm{k}}|} \notag \\
  &\hspace{-5mm} \times \Biggl[ \biggl( \frac{\partial \varepsilon_{\bm{k}}}{\partial k_{y}} \biggr) \biggl\{ D_{14} (\bm{k}) + D_{23} (\bm{k}) \biggr\} \biggl\{ T^{2}_{\bm{k} -} + T^{2}_{\bm{k} +} - 1 \biggr\} \notag \\
    & \hspace{-0mm} + \sin k_{y} \hspace{0.7mm} \biggl\{ D_{14} (\bm{k}) T^{(1)}_{AB} (\bm{k}) + D_{23} (\bm{k}) T^{(1)}_{BA} (\bm{k}) \biggr\} \notag \\
    & \hspace{-0mm} \times \biggl\{ T_{\bm{k} -} T_{\bm{k} +} + \sqrt{1 - T^{2}_{\bm{k} -}} \sqrt{1 - T^{2}_{\bm{k} +}}  \biggr\} \Biggr] , \label{SHC_yx_C} \\
  \sigma^{{\rm SHE}(\rm NC)}_{yx} &= \frac{\alpha g \mu_{\rm{B}} e}{N} \sum_{\bm{k}} \frac{\cos k_{x} \sin k_{y}}{|\bm{g}_{\bm{k}}|} \notag \\
  &\hspace{-5mm} \times \Biggl[ \biggl( \frac{\partial \varepsilon_{\bm{k}}}{\partial k_{y}} \biggr) \biggl\{ D_{12} (\bm{k}) - D_{34} (\bm{k}) \biggr\} \biggl\{ T^{2}_{\bm{k} -} - T^{2}_{\bm{k} +} \biggr\} \notag \\
    & \hspace{-0mm} + \sin k_{y} \hspace{0.7mm} \biggl\{ D_{12} (\bm{k}) T^{(2)}_{AB} (\bm{k}) - D_{34} (\bm{k}) T^{(2)}_{BA} (\bm{k}) \biggr\} \notag \\
    & \hspace{-0mm} \times \biggl\{ T_{\bm{k} -} \sqrt{1 - T^{2}_{\bm{k} +}} - T_{\bm{k} +} \sqrt{1 - T^{2}_{\bm{k} -}} \biggr\} \Biggr] . \label{SHC_yx_NC}
\end{align}
We have introduced $T^{(1)}_{ll'} (\bm{k})$, $T^{(2)}_{ll'} (\bm{k})$, and $D_{ab} (\bm{k})$ by
\begin{align}
  T^{(1)}_{ll'} (\bm{k}) &= \Delta_{l} T_{\bm{k} -} T_{\bm{k} +} - \Delta_{l'}  \sqrt{1 - T^{2}_{\bm{k} -}} \sqrt{1 - T^{2}_{\bm{k} +}}, \label{T1_ll'} \\
  T^{(2)}_{ll'} (\bm{k}) &= \Delta_{l} T_{\bm{k} -} \sqrt{1 - T^{2}_{\bm{k} +}} + \Delta_{l'} T_{\bm{k} +} \sqrt{1 - T^{2}_{\bm{k} -}}, \label{T2_ll'}
\end{align}
and
\begin{align}
  D_{ab} (\bm{k}) &= \int^{\infty}_{-\infty} \frac{d \varepsilon}{\pi} f (\varepsilon) & \notag \\
  & \hspace{-10.0mm} \times \Biggl[ - \frac{\partial {\rm Re} G^{\rm{R}}_{\bm{k} b} (\varepsilon) }{\partial \varepsilon} \hspace{0.5mm} {\rm Im} G^{\rm{R}}_{\bm{k} a} (\varepsilon) \hspace{0.7mm} + \hspace{0.7mm} {\rm Im} G^{\rm{R}}_{\bm{k} b} (\varepsilon) \hspace{0.5mm} \frac{\partial {\rm Re} G^{\rm{R}}_{\bm{k} a} (\varepsilon) }{\partial \varepsilon} \Biggr] . \label{D_ab}
\end{align}
From Eq.~(\ref{D_ab}), we have $D_{ab} (\bm{k})=0$ when $E_{\bm{k} a} = E_{\bm{k} b}$.
Thus, the spin Hall conductivity comes from the interband contribution. 
Using the retarded Green function in the band basis 
$G^{\rm{R}}_{\bm{k} a} (\varepsilon) = \left(\varepsilon - E_{\bm{k} a} + i \gamma\right)^{-1}$, 
we obtain 
\begin{equation}
  D_{ab} (\bm{k}) = - \frac{f (E_{\bm{k} a}) - f (E_{\bm{k} b})}{\bigl(E_{\bm{k} a} - E_{\bm{k} b}\bigr)^{2}}, \label{D_ab_gamma_zero}
\end{equation}
for $E_{\bm{k} a} \ne E_{\bm{k} b}$ and $\gamma \rightarrow +0$. 

\subsection{Tetragonal system}
\begin{center}
\begin{figure}[htbp]
 \begin{center}
   \includegraphics[width=0.75\hsize]{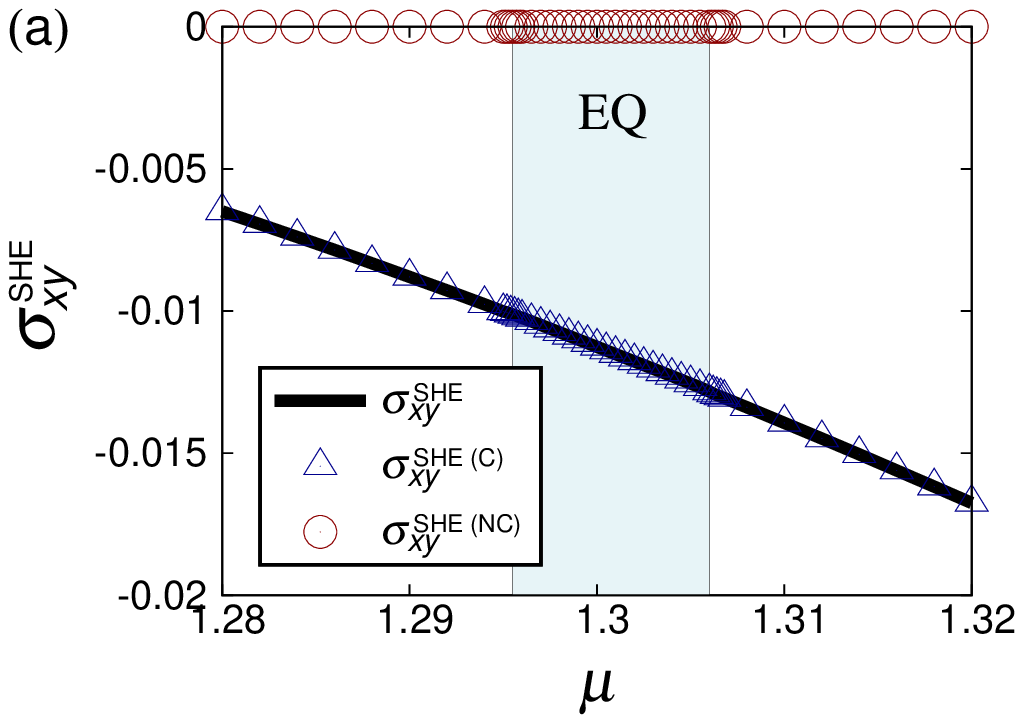}
\\
\vspace{5mm}
   \includegraphics[width=0.75\hsize]{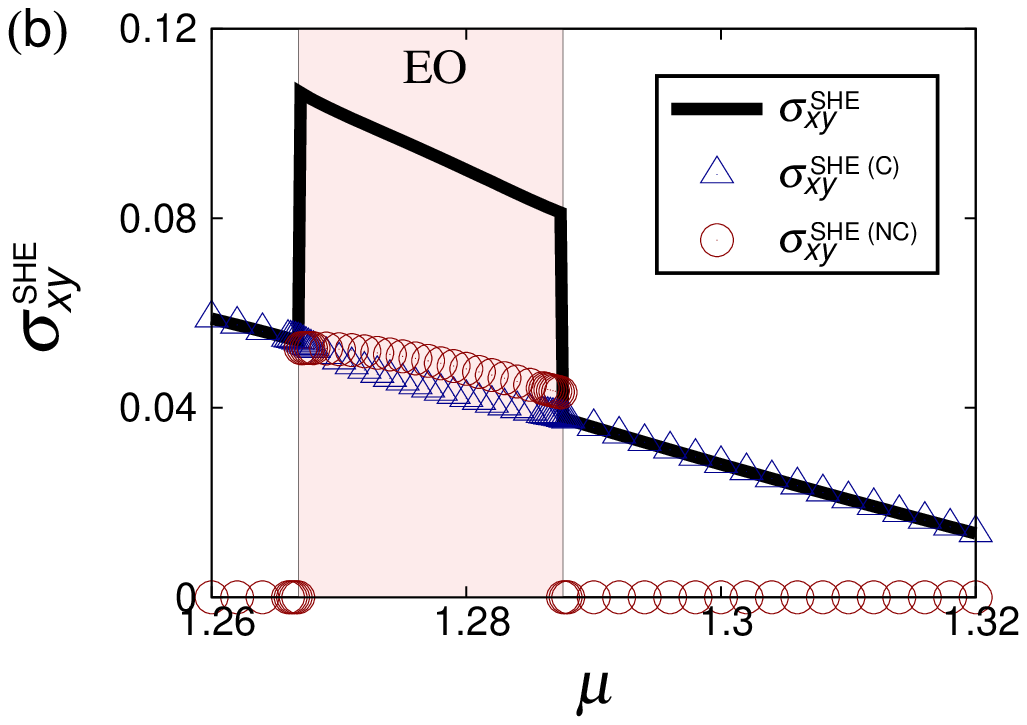}
   \caption{(Color online) Chemical potential dependence of the spin Hall conductivity.
     (a) The EQ state is stabilized for $g_{1} = 0.45$, $g_{2} = 0.05$, $\alpha = 0.1$ (blue shaded region).
     (b) The EO state is stabilized for  $g_{1} = 0.45$, $g_{2} = -0.05$, $\alpha = 0.35$ (red shaded region).
     The black solid lines, blue triangles, and red circles show
     $\sigma^{\rm{SHE}}_{xy}$, $\sigma^{{\rm SHE}(\rm {C})}_{xy}$, and $\sigma^{{\rm SHE}(\rm {NC})}_{xy}$ at $T=0.01$, respectively.
     Note that $\sigma^{{\rm SHE}(\rm {NC})}_{xy}$ is finite only in the EO state.
   }
   \label{SHC_cp}
 \end{center}
\end{figure}
\end{center}

In this subsection, we consider the tetragonal system and set $\delta t_{1} = 0$, $t_2=0.35$. 
Here we compare the spin Hall effect in the EQ and EO states. 
First, Fig.~\ref{SHC_cp} shows the spin Hall conductivity $\sigma^{\rm{SHE}}_{xy}$, $\sigma^{{\rm SHE}(\rm{C})}_{xy}$, 
and $\sigma^{{\rm SHE}({\rm NC})}_{xy}$ as a function of the chemical potential.
When we assume a parameter set $(\alpha, g_1, g_2)=(0.1, 0.45,0.05)$, the EQ state is stable at around $\mu =1.3$.  
On the other hand, the EO state is stable at $1.2669 < \mu < 1.2875$ for $(\alpha, g_1, g_2)=(0.35, 0.45,-0.05)$. 
Note that in a large ASOC region, sign of $g_2$ mainly determines the stability of the EQ and EO states~\cite{Hitomi_EO_JPSJ2016}. 
In normal and EQ states, $\sigma^{\rm{SHE}}_{xy} = \sigma^{{\rm SHE}(\rm {C})}_{xy}$ 
since $\sigma^{{\rm SHE}(\rm {NC})}_{xy}$ vanishes. 
Therefore, an effect of EQ order on the spin Hall conductivity is not visible in Fig.~\ref{SHC_cp}(a).
This is because the inversion symmetry is preserved in the EQ state.

On the other hand, the spin Hall conductivity shows a pronounced feature in the EO state.
As shown by Fig.~\ref{SHC_cp}(b), not only $\sigma^{{\rm SHE}(\rm {C})}_{xy}$ but also 
$\sigma^{{\rm SHE}(\rm {NC})}_{xy}$ are finite in the EO state, and the magnitude of the two terms is comparable. 
Therefore, the spin Hall conductivity $\sigma^{\rm{SHE}}_{xy}$ is almost doubled in the EO state 
with a discontinuous jump at the phase boundaries.

\begin{figure}[htbp]
 \begin{center}
   \includegraphics[width=0.90\hsize]{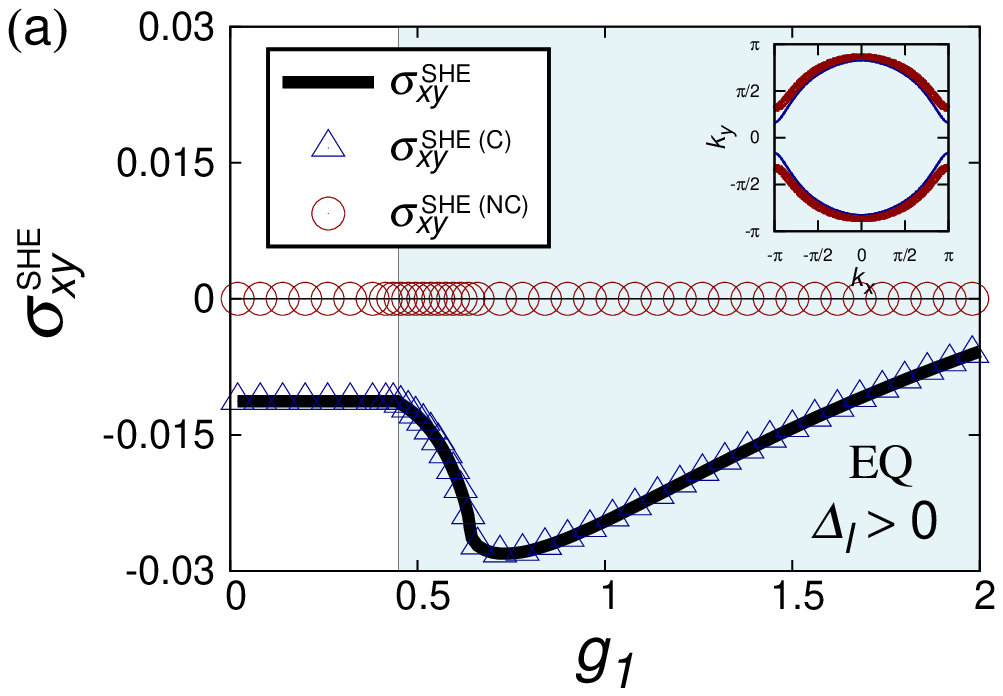}
\\
\vspace{5mm}
\hspace{-15mm}
   \includegraphics[width=0.75\hsize]{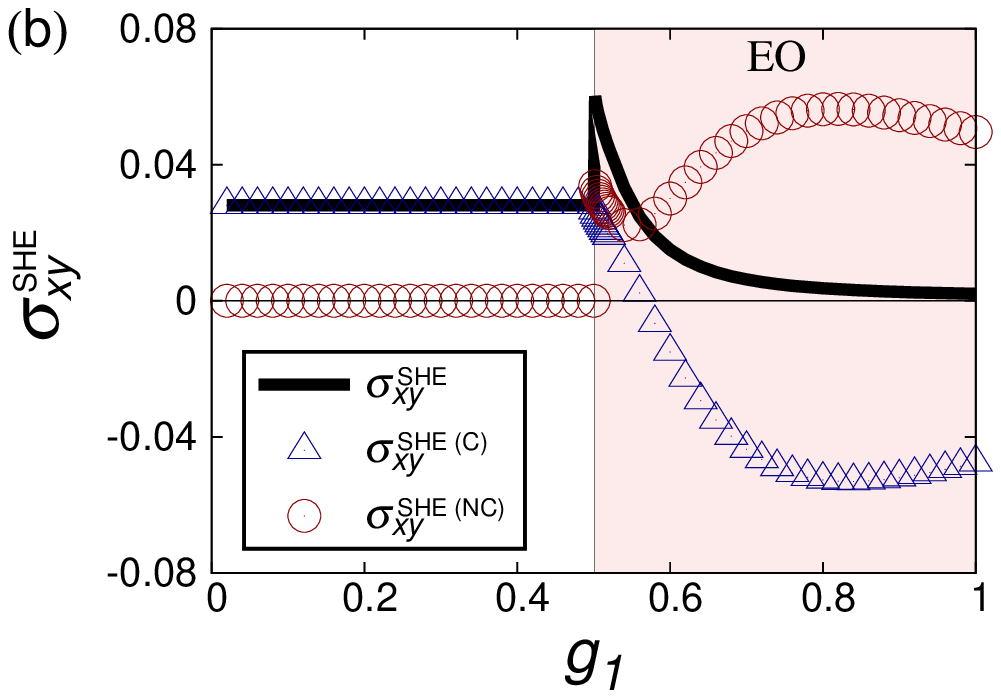}
   \caption{(Color online) Forward scattering interaction dependence of the spin Hall conductivity.
     In (a) we assume $g_{2} = 0.05$, $\alpha = 0.1$, and $\mu = 1.3$ to discuss the EQ phase. 
     Since the EQ state breaks $C_4$ rotation symmetry, $\sigma^{\rm{SHE}}_{xy}$ depends on sign of $\Delta_{A}=\Delta_{B}$. 
     We show the result for $\Delta_{A}=\Delta_{B} >0$. The other case shows the qualitatively same behaviors.  
     The inset shows the Fermi surface at $g_1=0.7$.
     In (b) we assume $g_{2} = -0.05$, $\alpha = 0.35$, and $\mu = 1.3$ to discuss the EO phase. The temperature is $T = 0.01$.
     The black solid lines, blue triangles, and red circles show
     $\sigma^{\rm{SHE}}_{xy}$, $\sigma^{{\rm SHE}(\rm {C})}_{xy}$, and $\sigma^{{\rm SHE}(\rm {NC})}_{xy}$, respectively. 
   }
   \label{SHC_g1}
 \end{center}
\end{figure}

Next, Fig.~\ref{SHC_g1} shows the spin Hall conductivity as a function of the forward scattering coupling constant $g_{1}$.
Figure~\ref{SHC_g1}(a) reveals that the spin Hall conductivity does not show discontinuity at the critical point of EQ order, 
consistent with Fig.~\ref{SHC_cp}(a). The spin Hall conductivity shows a peak around the Lifshitz transition point 
where the Fermi surface changes the topology [see the inset of Fig.~\ref{SHC_cp}(a)].
On the other hand, in Fig.~\ref{SHC_g1}(b), $\sigma^{{\rm SHE}}_{xy}$ shows a jump at the critical point of the EO state, 
consistent with Fig.~\ref{SHC_cp}(b). 
However, the spin Hall conductivity is suppressed by further increasing $g_1 > 0.6$, owing to 
cancellation of $\sigma^{{\rm SHE}(\rm {C})}_{xy}$ and $\sigma^{{\rm SHE}(\rm {NC})}_{xy}$.
Let us comment that the spin Hall conductivity is sensitive to the topology of Fermi surfaces.
For instance, the sign change of $\sigma^{{\rm SHE}(\rm {C})}_{xy}$ in Fig.~\ref{SHC_g1}(b) coincides the Lifshitz transition 
where one of the electron Fermi surfaces changes to hole-like. The sign of $\sigma^{{\rm SHE}(\rm {C})}_{xy}$ 
in the normal state is different between Figs.~\ref{SHC_g1}(a) and \ref{SHC_g1}(b) because the Fermi surfaces have different topology, that is, 
electron Fermi surfaces appear in the latter.

\begin{figure}[htbp]
 \begin{center}
   \includegraphics[width=0.70\hsize]{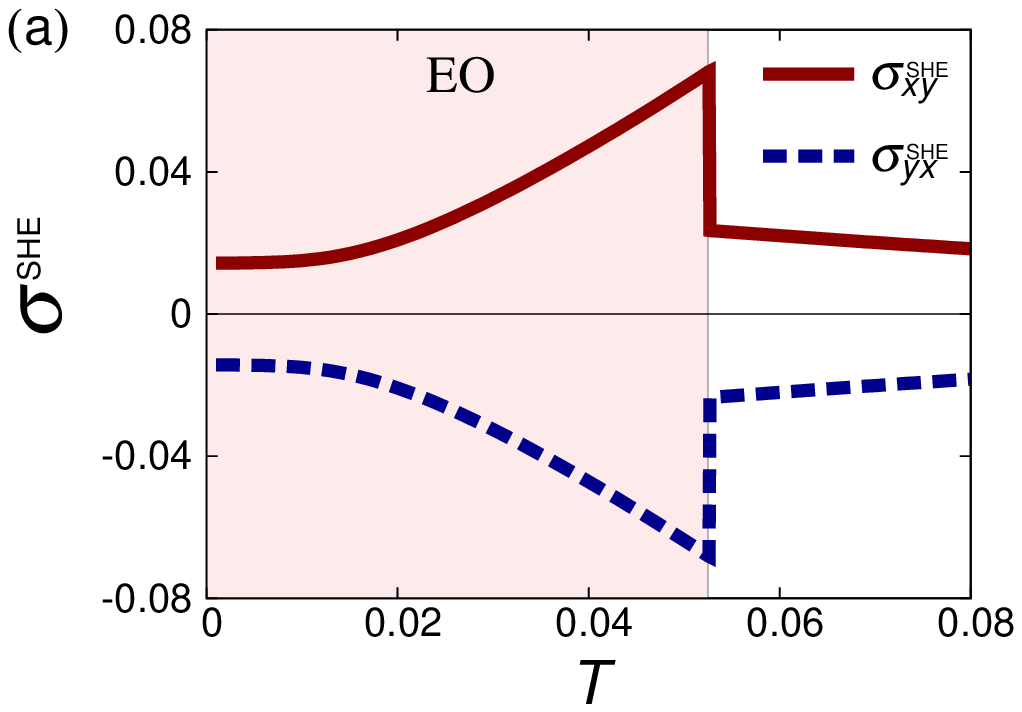}
\\
\vspace{5mm}
   \includegraphics[width=0.70\hsize]{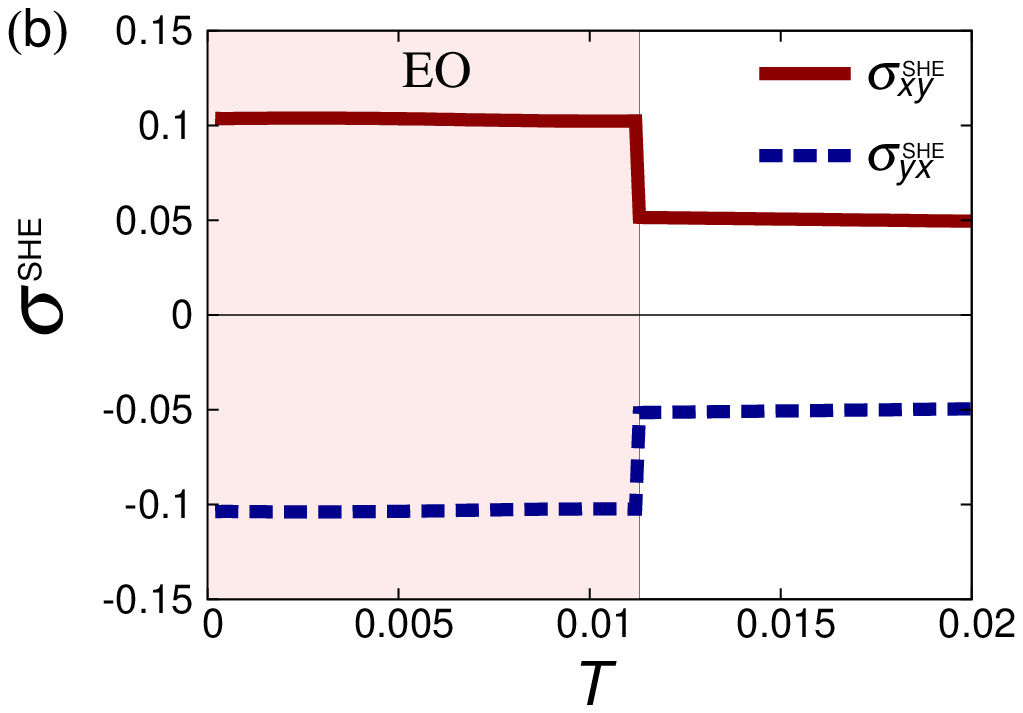}
\\
\vspace{5mm}
   \includegraphics[width=0.70\hsize]{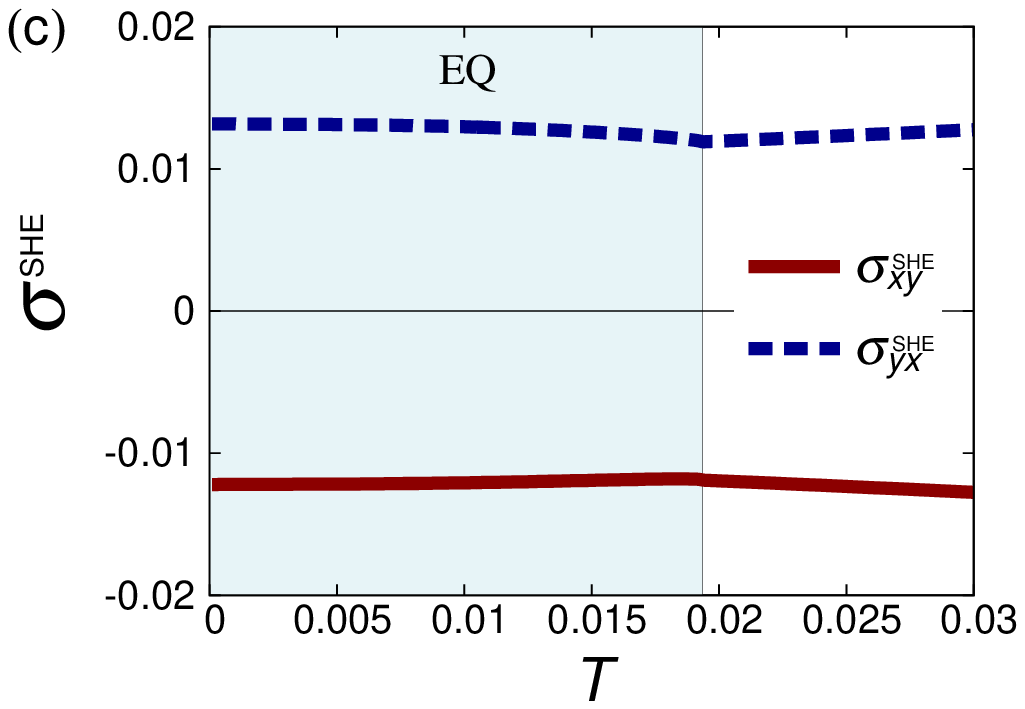}
   \caption{(Color online) Temperature dependence of the spin Hall conductivity.
     (a) $g_{1} = 0.6$, $g_{2} = -0.05$, $\alpha = 0.35$, and $\mu = 1.3$.
     (b) $g_{1} = 0.45$, $g_{2} = -0.05$, $\alpha = 0.35$, and $\mu = 1.27$.
     (c) $g_{1} = 0.5$, $g_{2} = 0.05$, $\alpha = 0.1$, and $\mu = 1.3$.
     The red solid lines and blue dashed lines show $\sigma^{\rm{SHE}}_{xy}$ and $\sigma^{\rm{SHE}}_{yx}$, respectively.
   }
   \label{SHC_tetragonal}
 \end{center}
\end{figure}

The temperature dependence of the spin Hall conductivity is plotted in Fig.~\ref{SHC_tetragonal}. 
Results for $g_{1} = 0.6$ and $g_{1} = 0.45$ are shown in Figs.~\ref{SHC_tetragonal}(a) and \ref{SHC_tetragonal}(b), respectively.
We see the jump in the spin Hall conductivity at the transition temperature of EO order in both parameters. 
This is a signature of parity violation, which can be tested by experiments. 
Although the spin Hall effect is suppressed at low temperatures for a large $g_1 =0.6$  
where $|\Delta_{l}| \gg \alpha |\bm{g}_{\bm{k}}|$, the spin Hall effect shows discontinuous enhancement 
at the phase boundary due to appearance of the $\sigma^{{\rm SHE}(\rm {NC})}_{\mu \nu}$ term. 
For a small $g_1 =0.45$ the spin Hall conductivity is nearly temperature independent below $T_{\rm c}$ 
since $|\Delta_{l}| < \alpha |\bm{g}_{\bm{k}}|$ at all temperatures. 
On the other hand, Fig.~\ref{SHC_tetragonal}(c) reveals that the spin Hall effect does not show a signature of EQ order. 
In our calculation, the spin Hall conductivity shows a discontinuous jump because we take $\gamma \rightarrow +0$. 
The discontinuity is smeared by a finite scattering rate $\gamma$. Fluctuation exchange approximation for the Hubbard model leads to 
$\gamma \sim 20$meV at $T=200$K,~\cite{Yanase_review} that is comparable to the spin-orbit splitting $\Delta_{\rm so}  \sim 10$meV for the parameters adopted in Sec.~3.3. 
Therefore, although a considerable smearing effect would occur, rapid increase in the spin Hall effect is expected in the EO state.

Finally, structure of the spin Hall conductivity tensor is discussed. 
We analytically confirmed $\sigma^{\rm{SHE}}_{xy} = - \sigma^{\rm{SHE}}_{yx}$ in the EO state and $\sigma^{\rm{SHE}}_{xx} = \sigma^{\rm{SHE}}_{yy} = 0$
(see Appendix A). 
The former relation has been shown in the numerical results (Fig.~\ref{SHC_tetragonal}).
Thus, the transverse spin current is induced by the electric field, namely, ${\bf J}^{sz} \perp {\bf E}$, irrespective of the direction of ${\bf E}$. 
This relation breaks down when the $C_4$ rotation symmetry is broken, as we will show in the next subsection.

\subsection{Orthorhombic system}
\begin{figure}[htbp]
 \begin{center}
   \includegraphics[width=0.70\hsize]{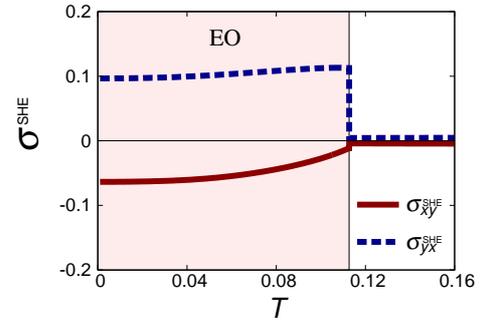}
   \caption{(Color online) (a) Temperature dependence of the spin Hall conductivity, 
     $\sigma^{\rm{SHE}}_{xy}$ (red solid line) and $\sigma^{\rm{SHE}}_{yx}$ (blue dashed line),
     in the orthorhombic system. 
     The parameters are the same as Sec. 3.
     The EO state is stable below $T_{\rm c} \sim 0.1127$ (red shaded region).
   }
   \label{SHC_orthorombic_weak}
 \end{center}
\end{figure}

Here, we discuss the spin Hall effect in the orthorhombic system, 
considering the pseudogap phase in the high-$T_{\rm c}$ cuprate superconductor YBCO. 
In this subsection, we adopt the same parameter set as that in Sec. 3.
Figure~\ref{SHC_orthorombic_weak} shows temperature dependence of $\sigma^{\rm{SHE}}_{xy}$ and $\sigma^{\rm{SHE}}_{yx}$, 
which show discontinuous enhancement below the transition temperature of EO order. 
In contrast to the tetragonal system, the relation $\sigma^{\rm{SHE}}_{xy} = -\sigma^{\rm{SHE}}_{yx}$ does not hold 
because the symmetry is reduced from $D_{2d}$ to $C_{2v}$. 

\begin{figure}[htbp]
 \begin{center}
\hspace{-10mm}
   \includegraphics[width=0.75\hsize]{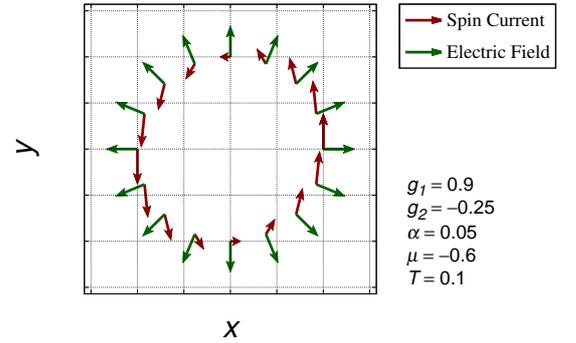}
   \caption{(Color online) Relation between the electric field direction and the spin current in the EO state at $T = 0.1$.
   The parameters are the same as Fig.~\ref{SHC_orthorombic_weak}. 
   The electric field is illustrated by green arrows and the corresponding spin current is plotted by red arrows. 
   }
   \label{SHC_orthorombic_direction}
 \end{center}
\end{figure}

Since we have proved  $\sigma^{\rm{SHE}}_{xx} = \sigma^{\rm{SHE}}_{yy} =0$, the spin current 
in various electric field directions is obtained from Fig.~\ref{SHC_orthorombic_weak}. 
The result is illustrated in Fig.~\ref{SHC_orthorombic_direction}. 
Because of  $\sigma^{\rm{SHE}}_{xy} \ne -\sigma^{\rm{SHE}}_{yx}$, the spin current is no longer perpendicular to 
the electric field. The longitudinal spin current is induced when the electric field is applied away from the [100] and [010] directions. The deviation from the orthogonal relation, ${\bf J}^{sz} \perp {\bf E}$, is enhanced for a larger $g_1$.

\section{Edelstein Effect}
In this section, we investigate spin polarization induced by electric current~\cite{
Edel'shtein_ME_PRL1995,Levitov_ME_JETP1985,Yip_SC_ME_PRB2002,Fujimoto_SC_ME_PRB2005,
Fujimoto_SC_ME_JPSJ2007_1,Fujimoto_SC_ME_JPSJ2007_2,Edelstein1990}, which was called kinetic magnetoelectric effect. 
This response is an origin of the spin-orbit torque for spintronics applications~\cite{Manchon2008,Garate2009,Miron2010}, 
and recently called Edelstein effect. 
Because the Edelstein effect occurs in spin-orbit coupled noncentrosymmetric metals,  
it is naturally expected that the Edelstein effect occurs in the EO state accompanied by spontaneous parity violation.

The Edelstein effect is represented by
\begin{equation}
  M_{\mu} = - \Upsilon_{\mu \nu} \hspace{0.5mm} E_{\nu} , \label{magnetoelectric_effect}
\end{equation}
where $M_{\mu}$ is the magnetic moment induced by the electric field $E_{\nu}$.
The magnetoelectric tensor $\Upsilon_{\mu \nu}$ is calculated by Kubo formula, 
\begin{equation}
  \Upsilon_{\mu \nu} = \lim_{\omega \rightarrow 0} \frac{K^{\rm{ME}}_{\mu \nu} (\omega) - K^{\rm{ME}}_{\mu \nu } (0)}{i \omega} , \label{MEC_kubo} 
\end{equation}
\begin{equation}
  K^{\rm{ME}}_{\mu \nu} (\omega) = \left. K^{\rm{ME}}_{\mu \nu} (i \omega_{n}) \right|_{i \omega_{n} \rightarrow \omega + i \hspace{0.5mm} 0} . \label{MEC_response_function_kaiseki}
\end{equation}
The correlation function is 
\begin{equation}
  K^{\rm{ME}}_{\mu \nu} (i \omega_{n}) = \frac{1}{N} \int^{1/T}_{0} d \tau \hspace{0.5mm} \langle T_{\tau} \{ S_{\mu} (\tau) \hspace{0.5mm} J_{\nu} (0) \} \rangle \hspace{0.5mm} e^{i \omega_{n} \tau} , \label{MEC_response_function}
\end{equation}
where $S_{\mu}$ are spin operators defined by
\begin{equation}
  S_{\mu} =  \frac{1}{2} g \mu_{\rm{B}} \sum_{\bm{k}} \hat{C}_{\bm{k}}^\dag \hspace{0.7mm}
  \scalebox{1.3}{$\displaystyle
    {\footnotesize
      \begin{pmatrix}
        \sigma^{\,\mu} & 0            \\
        0           & \sigma^{\,\mu}
      \end{pmatrix}
    } $}
  \hspace{0.7mm} \hat{C}_{\bm{k}}. \label{Spin_matrix} 
\end{equation}
Carrying out the unitary transformation, we obtain the band representation of 
spin operators $S_{\mu}$ and charge current operators $J_{\nu}$.
Then, the magnetoelectric tensor $\Upsilon_{\mu \nu}$ is decomposed into the three terms 
\begin{equation}
  \Upsilon_{\mu \nu} = \Upsilon^{\rm intra}_{\mu \nu} + \Upsilon^{{\rm inter}(\rm C)}_{\mu \nu} + \Upsilon^{{\rm inter}(\rm NC)}_{\mu \nu}. \label{MEC_intra_and_C_and_NC}
\end{equation}
The first term $\Upsilon^{\rm intra}_{\mu \nu}$ is the intraband term, while the others are the interband terms.
As we did for the spin Hall conductivity [Eq.~(\ref{SHC_C_and_NC})],
we separate the interband contribution into the two terms 
$\Upsilon^{{\rm inter}(\rm C)}_{\mu \nu}$ and $\Upsilon^{{\rm inter}(\rm NC)}_{\mu \nu}$.

Let us show the analytic expression of $\Upsilon_{xy}$, 
\begin{align}
  \Upsilon^{\rm intra}_{xy} &= \frac{g \mu_{\rm{B}} e}{2 N} \sum_{\bm{k}} \frac{\sin k_{y}}{|\bm{g}_{\bm{k}}|} \notag \\
  &\hspace{0.0mm} \times \Biggl[ \biggl( \frac{\partial E_{\bm{k} 1}}{\partial k_{y}} \biggr) \hspace{0.7mm} I_{11} (\bm{k}) - \biggl( \frac{\partial E_{\bm{k} 2}}{\partial k_{y}} \biggr) \hspace{0.7mm} I_{22} (\bm{k}) \notag \\
  &\hspace{0.0mm} + \biggl( \frac{\partial E_{\bm{k} 3}}{\partial k_{y}} \biggr) \hspace{0.7mm} I_{33} (\bm{k}) - \biggl( \frac{\partial E_{\bm{k} 4}}{\partial k_{y}} \biggr) \hspace{0.7mm} I_{44} (\bm{k}) \Biggr] , \label{MEC_xy_intra} \\
  \Upsilon^{{\rm inter}(\rm C)}_{xy} &= \frac{\alpha g \mu_{\rm{B}} e}{N} \sum_{\bm{k}} \frac{\sin^{2} k_{x} \cos k_{y}}{|\bm{g}_{\bm{k}}|^{2}} \notag \\
  &\hspace{0.0mm} \times \biggl[ - I_{41} (\bm{k}) + I_{32} (\bm{k}) \biggr] \hspace{0.5mm} \biggl[ T^{2}_{\bm{k} -} + T^{2}_{\bm{k} +} - 1 \biggr], \label{MEC_xy_C} \\
  \Upsilon^{{\rm inter}(\rm NC)}_{xy} &= \frac{\alpha g \mu_{\rm{B}} e}{N} \sum_{\bm{k}} \frac{\sin^{2} k_{x} \cos k_{y}}{|\bm{g}_{\bm{k}}|^{2}} \notag \\
  &\hspace{0.0mm} \times \biggl[ - I_{21} (\bm{k}) + I_{43} (\bm{k}) \biggr] \hspace{0.5mm} \biggl[ T^{2}_{\bm{k} -} - T^{2}_{\bm{k} +} \biggr]. \label{MEC_xy_NC}
\end{align}
%
%
We have introduced $I_{ab} (\bm{k})$ by
\begin{equation}
  I_{ab} (\bm{k}) = \int^{\infty}_{-\infty} \frac{d \varepsilon}{\pi} \{ - f' (\varepsilon) \}  \hspace{1mm} {\rm Im} G^{\rm{R}}_{\bm{k} a} (\varepsilon) \hspace{0.7mm} {\rm Im} G^{\rm{R}}_{\bm{k} b} (\varepsilon). \label{I_ab}
\end{equation}
Similarly, $\Upsilon^{\rm intra}_{yx}$ is obtained by exchanging $k_x \leftrightarrow k_y$ and adding the negative sign. 
As we will show later, the intraband term $\Upsilon^{\rm intra}_{\mu \nu}$ is dominant.

For simplicity, we here assume that the quasiparticle's damping $\gamma$ is independent of the band index and momentum.
Then, the intraband term is reduced to
\begin{align}
  \Upsilon^{\rm intra}_{xy} &= \frac{g \mu_{\rm{B}} e}{4 \gamma N} \sum_{\bm{k}} \frac{\sin k_{y}}{|\bm{g}_{\bm{k}}|} \notag \\
  &\hspace{0.0mm} \times \Biggl[ \biggl( \frac{\partial E_{\bm{k} 1}}{\partial k_{y}} \biggr) \hspace{0.7mm} \left\{ - f' (E_{\bm{k} 1}) \right\} - \hspace{1mm} \biggl( \frac{\partial E_{\bm{k} 2}}{\partial k_{y}} \biggr) \hspace{0.7mm} \left\{ - f' (E_{\bm{k} 2}) \right\} \notag \\
  &\hspace{0.0mm} + \biggl( \frac{\partial E_{\bm{k} 3}}{\partial k_{y}} \biggr) \hspace{0.7mm} \left\{ - f' (E_{\bm{k} 3}) \right\} \hspace{1mm} - \hspace{1mm} \biggl( \frac{\partial E_{\bm{k} 4}}{\partial k_{y}} \biggr) \hspace{0.7mm} \left\{ - f' (E_{\bm{k} 4}) \right\} \Biggr], \label{MEC_xy_intra_gamma}
\end{align}
by using $G^{\rm{R}}_{\bm{k} a}(\varepsilon) = (\varepsilon - E_{{\bf k}a} + i \gamma)^{-1}$. 
In the following we numerically calculate the other terms $\Upsilon^{{\rm inter}(\rm C)}_{\mu \nu}$ and $\Upsilon^{{\rm inter}(\rm NC)}_{\mu \nu}$.

\subsection{Tetragonal system}
\begin{figure}[htbp]
 \begin{center}
   \includegraphics[width=0.85\hsize]{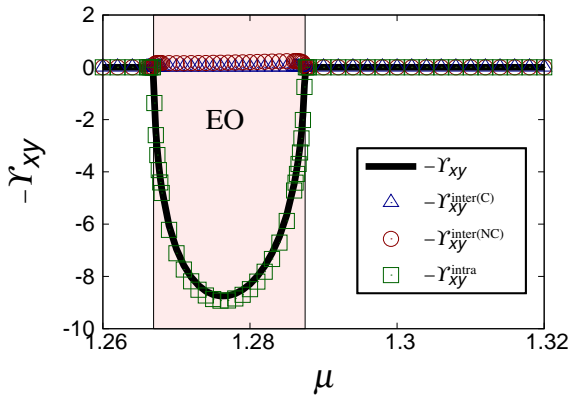}
   \caption{(Color online) The magnetoelectric coefficient $\Upsilon_{xy}$ as a function of the chemical potential
     for $g_{1} = 0.45$, $g_{2} = -0.05$, $\alpha = 0.35$, $T = 0.01$ and $\gamma = 1.0 \times 10^{-3}$.
     The black solid line, blue triangles, red circles, and green squares show
     $-\Upsilon_{xy}$, $-\Upsilon^{{\rm inter}(\rm {C})}_{xy}$, $-\Upsilon^{{\rm inter}(\rm {NC})}_{xy}$, and $-\Upsilon^{{\rm intra}}_{xy}$, respectively.
     The magnetoelectric coefficient is finite only in the EO state. 
     The dominant contribution comes from the intraband contribution $-\Upsilon^{{\rm intra}}_{xy}$.
   }
   \label{MEC_tetragonal_cp}
 \end{center}
\end{figure}

Now we calculate the Edelstein effect in the tetragonal system by setting $\delta t_{1} = 0$.
Figure~\ref{MEC_tetragonal_cp} shows the chemical potential dependence of 
the magnetoelectric coefficient $\Upsilon_{xy}$. 
In contrast to the spin Hall effect, the Edelstein effect does not occur in centrosymmetric systems. 
Therefore, $\Upsilon_{x y}$ disappears in the normal state. On the other hand, a finite 
$\Upsilon_{xy}$ is obtained in the EO state. Thus, the Edelstein effect is a signature of parity-violating electric order.~\cite{Watanabe2018a} 

It is shown that $\Upsilon_{xy}$ is mainly given by the intraband contribution $\Upsilon^{\rm intra}_{xy}$, 
i.e., $|\Upsilon^{\rm intra}_{xy}| \gg |\Upsilon^{{\rm inter}(\rm {C})}_{xy} + \Upsilon^{{\rm inter}(\rm {NC})}_{xy}|$.
This is because the Edelstein effect is caused by the shift of Fermi surfaces under the electric current. 
We confirmed that the intraband contribution is dominant even for a large scattering rate, e.g., $\gamma = 0.1$.

\begin{figure}[htbp]
 \begin{center}
\hspace{-15mm}
   \includegraphics[width=0.70\hsize]{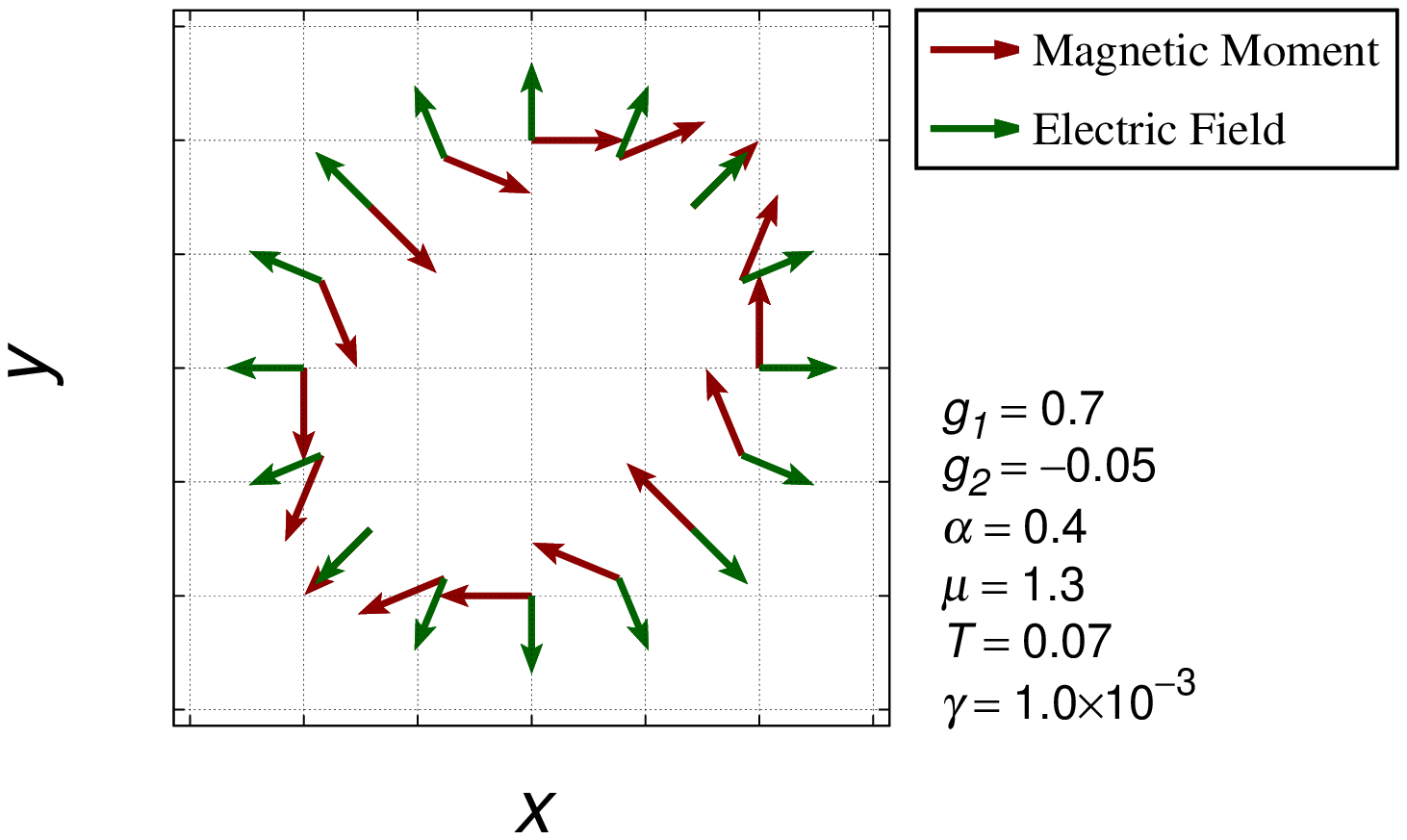}
   \caption{(Color online) Red arrows show the magnetic moment induced by the electric field depicted by green arrows. 
     We assume a tetragonal system.
   }
   \label{MEC_vect_tetragonal}
 \end{center}
\end{figure}

In contrast to the magnetoelectric tensor $\Upsilon_{xy} = -\Upsilon_{yx}$ in Rashba systems, 
$\Upsilon_{xy} = \Upsilon_{yx}$ holds and others are zero. 
Such magnetoelectric tensor has been shown by a group-theoretical analysis,~\cite{Watanabe2018a} 
although the $x$ axis is rotated by 45 degrees. 
The difference from Rashba systems is intuitively understood by considering the spin texture 
in the momentum space; $k_y \hat{x} + k_x \hat{y}$ for $D_{2d}$ symmetry while 
$-k_y \hat{x} + k_x \hat{y}$ for $C_{4v}$ symmetry.~\cite{Watanabe2018a} 
The non-polar EO state corresponds to the former, and the polar Rashba system corresponds to the latter. 
Reflecting the characteristic spin texture due to $D_{2d}$ point group symmetry, 
the EO state shows a characteristic electric-field-angle dependence of the magnetic moment in Fig.~\ref{MEC_vect_tetragonal}.
The longitudinal magnetoelectric response, ${\bf M} \parallel {\bf E}$, occurs for ${\bf E} \parallel$[110], 
while the transverse response, ${\bf M} \perp {\bf E}$, occurs for ${\bf E} \parallel$[100]. 
This unique field-angle dependence is distinct from Rashba systems and chiral crystals; 
${\bf M} \perp {\bf E}$ in polar Rashba systems~\cite{
Edel'shtein_ME_PRL1995,Levitov_ME_JETP1985,Yip_SC_ME_PRB2002,Fujimoto_SC_ME_PRB2005,
Fujimoto_SC_ME_JPSJ2007_1,Fujimoto_SC_ME_JPSJ2007_2,Edelstein1990},  
while ${\bf M} \parallel {\bf E}$ in chiral crystals~\cite{Yoda2015} lacking mirror symmetry. 
Thus, the Edelstein effect can distinguish the symmetry of  electron systems.

The Edelstein effect of $D_{2d}$-type may be observed in the spin-orbit coupled metallic state of Cd$_2$Re$_2$O$_7$. 
Below the structural transition temperature $T_{\rm s}=200$K, the crystal symmetry is 
reduced from cubic $O_h$ to tetragonal $D_{2d}$~\cite{Hiroi_Cd2Re2O7}. 
From the viewpoint of symmetry, the transition has been identified as electric dotriacontapole order~\cite{Watanabe2018a} 
or electric toroidal quadrupole order.~\cite{Hayami2018b} 
Although the electronic state and relevant multipole moment are different from our bilayer model, symmetry of the low temperature phase is the same as the EO state. Therefore, qualitatively the same magnetoelectric transport is expected.

\subsection{Orthorhombic system}
We turn to the orthorhombic system, and set $\delta t_{1} = 0.05$.
Symmetry of the EO state belongs to the noncentrosymmetric and polar $C_{2v}$ point group, and therefore, 
structure of the magnetoelectric tensor is different from the tetragonal EO state with $D_{2d}$ symmetry.

\begin{figure}[htbp]
 \begin{center}
   \includegraphics[width=0.70\hsize]{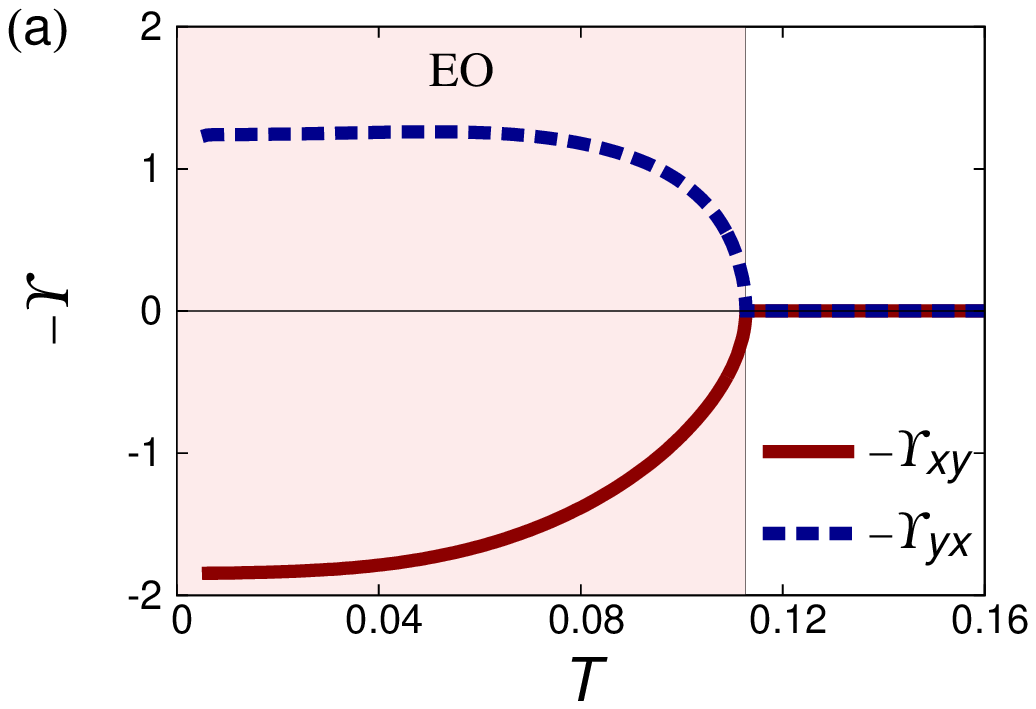}
\\
\vspace{5mm}
\hspace{-15mm}
   \includegraphics[width=0.70\hsize]{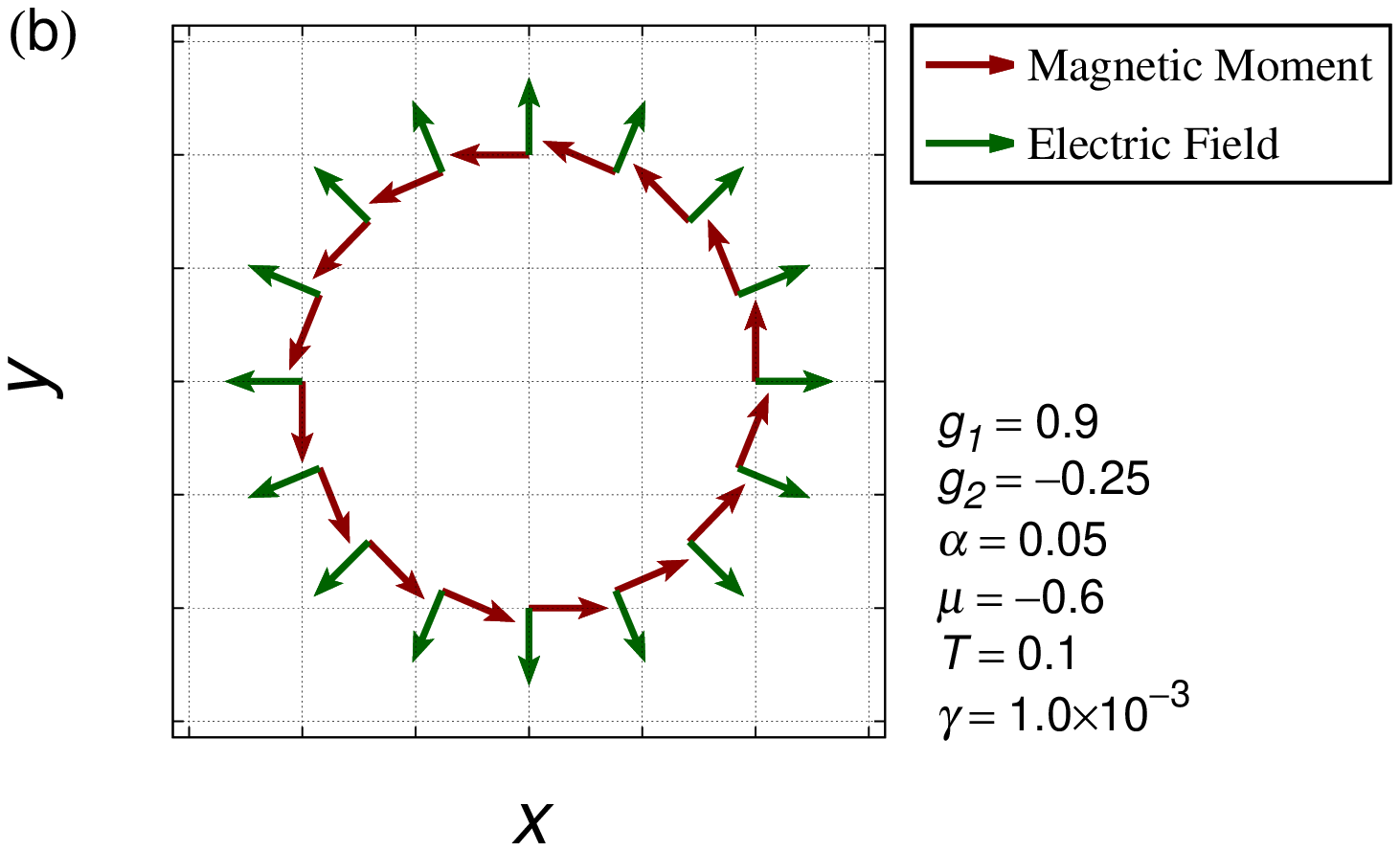}
   \caption{(Color online) (a) Temperature dependence of magnetoelectric coefficients for $\gamma = 1.0 \times 10^{-3}$.
     The other parameters are the same as Sec.~3. 
     The red solid line and blue dashed line show $-\Upsilon_{xy}$ and $-\Upsilon_{yx}$, respectively.
     (b) Electric-field-angle dependence of the magnetic moment at $T = 0.1$.
     The red arrows and green arrows illustrate the magnetic moment and the electric field, respectively.
   }
   \label{MEC_orthorombic_weak}
 \end{center}
\end{figure}

First, we show the results for $g_{1} = 0.9$ and $g_{2} = -0.25$ in Fig.~\ref{MEC_orthorombic_weak}.
The temperature dependence of $\Upsilon_{\mu\nu}$ reveals appearance of the Edelstein effect in the EO state. 
In contrast to the tetragonal system, the relation $\Upsilon_{xy} = \Upsilon_{yx}$ does not hold in the orthorhombic EO state. 
Moreover, signs of $-\Upsilon_{xy}$ and $-\Upsilon_{yx}$ are opposite. 
This structure of the magnetoelectric tensor results in nearly transverse magnetoelectric effect, ${\bf M} \perp {\bf E}$, 
shown in Fig.~\ref{MEC_orthorombic_weak}(b).  
Then, the field-angle dependence is similar to the Rashba system with $C_{4v}$ point group symmetry.

\begin{figure}[htbp]
 \begin{center}
   \includegraphics[width=0.70\hsize]{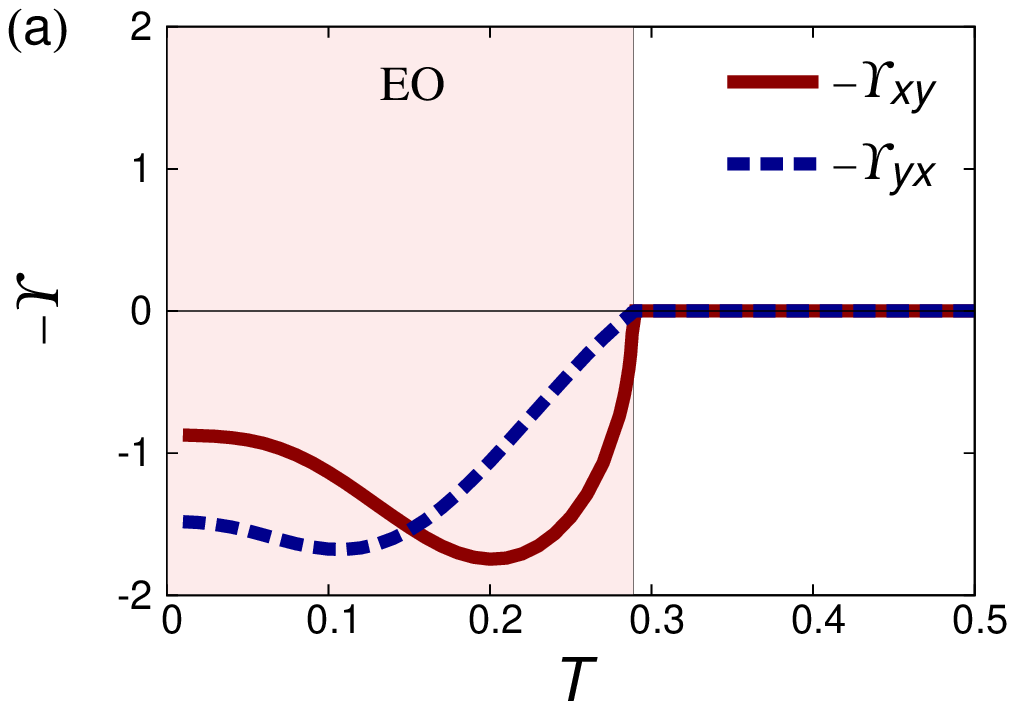}
\\
\vspace{5mm}
\hspace{-15mm}
   \includegraphics[width=0.70\hsize]{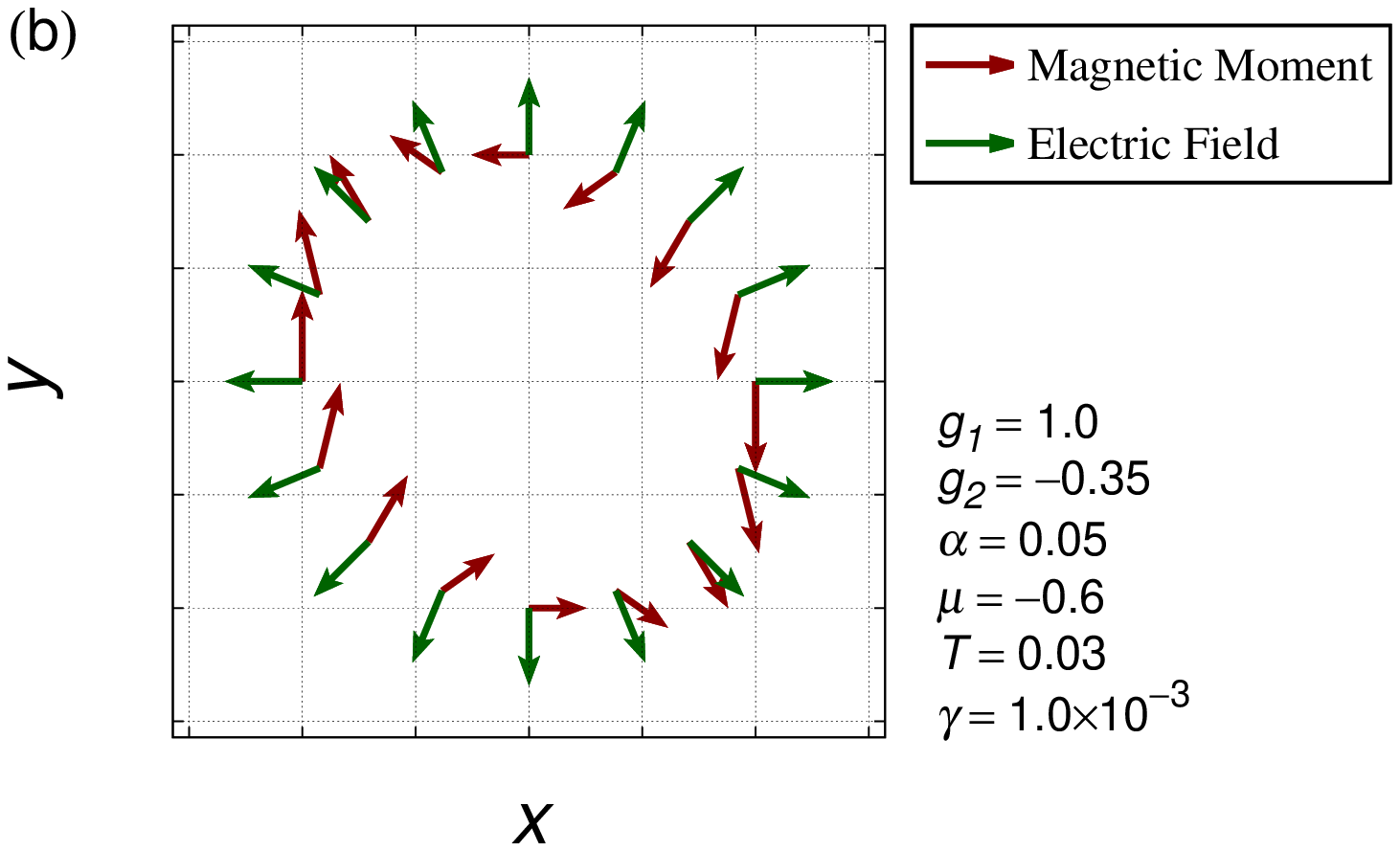}
   \caption{(Color online) (a) Temperature dependence of magnetoelectric coefficients
     for $g_{1} = 1.0$, $g_{2} = -0.35$. The other parameters are the same as Fig.~\ref{MEC_orthorombic_weak}. 
     (b) Electric-field-angle dependence of the magnetic moment at $T = 0.03$.
   }
   \label{MEC_orthorombic_strong}
 \end{center}
\end{figure}

The field-angle dependence significantly changes when the forward scattering interaction is increased. 
Figure~\ref{MEC_orthorombic_strong} obtained for $g_{1} = 1.0$ and $g_{2} = -0.35$ shows $\Upsilon_{xy} \sim \Upsilon_{yx}$.  
Thus, the Edelstein effect is similar to that in the EO state of tetragonal systems. 
The electric-field-angle dependence plotted in Fig.~\ref{MEC_orthorombic_strong}(b) actually resembles Fig.~\ref{MEC_vect_tetragonal}. 

From these results, it is implied that the Edelstein effect changes from the $C_{4v}$-type to the $D_{2d}$-type 
with increasing the EO moment. This crossover is allowed by symmetry, because the $C_{2v}$ point group 
of the orthorhombic EO state is a subgroup of $C_{4v}$ and $D_{2d}$. 
Intuitively speaking, a large EO moment with $D_{2d}$ symmetry causes the Edelstein effect of the $D_{2d}$-type. 
In our model, the crossover is accompanied by the Lifshitz transition of Fermi surfaces. 
For $g_{1} = 0.9$ an orthorhombicity by $\delta t_1 =0.05$ leads to some open Fermi surfaces, 
while only closed (two electronic- and two hole-)Fermi surfaces are obtained for $g_{1} = 1.0$ owing to a large EO moment $\Delta_{\rm EO}$. 
In the latter case, where $\Delta_{\rm EO} \gg \delta t_1$, nearly fourfold symmetric Fermi surfaces resemble those in $D_{2d}$ systems.
Accordingly, the Edelstein effect is almost $D_{2d}$-type.

\section{Summary and Discussions}
In this paper, we proposed a possibility of odd-parity EO order in bilayer high-$T_{\rm c}$ cuprate superconductors. 
Analyzing the forward scattering interaction in the bilayer Rashba model, we examined recent magnetic torque measurements for YBCO.~\cite{Sato_YBCO}
It is essential to take into account weak $C_4$ rotation symmetry breaking in the crystal structure of YBCO by CuO chains. 
In the orthorhombic crystals, the nematic EQ order is not a phase transition but a crossover. 
On the other hand, EO order with inversion symmetry breaking is a second order phase transition, consistent with experiments.~\cite{Sato_YBCO,Shekhter2013} 
The kink in temperature dependence of the magnetic torque~\cite{Sato_YBCO} is reproduced 
by our calculation for the EO state, although the nematic order leads to incompatible critical behaviors. 
Note that an alternative explanation without assuming phase transition has been proposed for the magnetic torque.~\cite{Morinari}  
Broken inversion symmetry by EO order is consistent with experimental observation by 
optical measurements.~\cite{Zhao2016} Combination of orthorhombic crystal distortion and EO order allows 
spontaneous electric polarization along the (001)-axis. 
Observation of polar lattice distortion may be an experimental test.  
In our calculation the critical temperature of EO order coincides with the onset of pseudogap phenomena. 
The spin susceptibility and density of states decrease below the critical temperature. 

The structure of the EO state is compatible with the crisscrossed stripe state~\cite{Maharaj_cuprate} 
which was proposed for the charge density wave (CDW) order observed below the pseudogap onset temperature.~\cite{Ghiringhelli_YBCO_CDW,Chang_YBCO_CDW,Comin2015} 
In the crisscrossed stripe state, the wave vector of CDW is orthogonal between bilayers.
Apparently, such CDW state is unstable in the nematic state which favors one of ${\bf q} \parallel [100]$ or ${\bf q} \parallel [010]$.
On the other hand, because of the opposite nematicity between bilayers, the crisscrossed stripe state may be stable in the EO state.

A key is time-reversal symmetry. Time-reversal symmetry is preserved in the EO state,  
although the time-reversal symmetry breaking magnetoelectric multipole~\cite{Lovesey2015,Fechner2016} and loop-current~\cite{Varma1997}  
have been proposed for the order parameter of pseudogap phase in cuprate superconductors. 
Polarized neutron scattering measurements~\cite{Fauque2006,Thro2015} and a polar Kerr effect measurement~\cite{Xia2008} 
supported broken time-reversal symmetry. 
However, these experiments reported different onset temperatures, and thus, interpretation of the data is still unsettled. 

For a future experimental test, we showed remarkably large in-plane anisotropy of spin susceptibility 
in the superconducting state. The anisotropy may be 100 times larger than that in the normal state. 
Such large anisotropy comes from the Van Vleck spin susceptibility. 
A NMR Knight shift measurement is proposed to detect highly anisotropic spin susceptibility. 

Furthermore, we investigated magnetoelectric transport properties in the EO state, 
which are caused by spontaneous parity violation. 
The spin Hall effect and the kinetic magnetoelectric effect (Edelstein effect) were calculated on the basis of the 
linear response theory. 
We divided the spin Hall conductivity into the two terms by Eq.~(\ref{SHC_C_and_NC}).
One is derived from the layer-dependent Rashba ASOC ($\sigma^{{\rm SHE}(\rm C)}_{\mu \nu}$),
and the other is given by the band splitting due to spontaneous parity violation in the EO state ($\sigma^{{\rm SHE}(\rm NC)}_{\mu \nu}$).
Owing to the appearance of the latter term, the spin Hall conductivity shows a discontinuous enhancement in the EO state. 
On the other hand, no noticeable change in the spin Hall conductivity is observed at the critical point of nematic (EQ) order. 
Thus, the EO state and nematic state can be distinguished by measuring the spin Hall effect. 
The Edelstein effect, namely, current-induced spin polarization occurs in the EO state, although it does not occur 
in the nematic state. This is because the Edelstein effect is prohibited in centrosymmetric systems. 
Measurements of Edelstein effect would give a constraint on symmetry of the pseudogap state.

The magnetoelectric transport in the EO states with tetragonal symmetry and orthorhombic symmetry have been compared. 
While the transverse spin Hall effect occurs in the tetragonal system, the longitudinal spin Hall conductivity appears in the orthorhombic system. 
The Edelstein effect in the tetragonal system shows a characteristic electric-field-angle dependence of $D_{2d}$-type, which is different from Rashba systems. 
On the other hand, the Edelstein effect in the orthorhombic system crossovers from 
$C_{4v}$-type to $D_{2d}$-type by increasing the forward scattering interaction strength. 
The $D_{2d}$-type Edelstein effect shows both transverse and longitudinal responses reflecting the non-polar and non-chiral $D_{2d}$ symmetry. In addition to the cuprate superconductors, we discussed Cd$_2$Re$_2$O$_7$.  
Observations of the magnetoelectric transport properties would demonstrate odd-parity multipole order 
and spontaneous space inversion symmetry breaking.

\begin{acknowledgement}
The authors are grateful to Y. Matsuda for fruitful discussions. 
This work was supported by Grant-in Aid for Scientific Research on Innovative Areas
``J-Physics" (JP15H05884) and ``Topological Materials Science" (JP18H04225) from JSPS of Japan, 
and by JSPS KAKENHI Grants No. JP15H05745, No. JP18H01178, and No. JP18H05227.
Numerical computation in this work was carried out at the Yukawa Institute Computer Facility.
T.H. was supported by a JSPS Fellowship for Young Scientists.
\end{acknowledgement}

%


\appendix

\section{Diagonal Spin Hall Conductivity and Magnetoelectric Coefficient}
We analytically show that the diagonal spin Hall conductivity $\sigma^{\rm{SHE}}_{\mu \mu}$ 
and magnetoelectric coefficient $\Upsilon_{\mu \mu} $ vanish, in accordance with a symmetry analysis. 
Calculating the matrix element of the spin operator in the band representation, we obtain the diagonal spin Hall conductivity
\begin{equation}
  \sigma^{\rm{SHE}}_{\mu \mu} = \sigma^{{\rm SHE}(\rm C)}_{\mu \mu} + \sigma^{{\rm SHE}(\rm NC)}_{\mu \mu}, \label{eq:APPENDIX_01}
\end{equation}
where we define 
\begin{equation}
  \sigma^{{\rm SHE}(\rm C)}_{\mu \mu} = \frac{1}{N} \sum_{\bm{k}} \sigma^{{\rm SHE}(\rm C)}_{\mu \mu} (\bm{k}), \label{eq:APPENDIX_02} \\
\end{equation}
\begin{align}
  \sigma^{{\rm SHE}(\rm C)}_{\mu \mu} (\bm{k}) &= \alpha g \mu_{\rm{B}} e \frac{\cos k_{\mu} \sin k_{\bar{\mu}}}{|\bm{g}_{\bm{k}}|}
\notag \\
& \hspace{-15mm} \times \Biggl[ T_{\bm{k} -} T_{\bm{k} +} + \sqrt{1 - T^{2}_{\bm{k} -}} \sqrt{1 - T^{2}_{\bm{k} +}} \Biggr] \notag \\
  &\hspace{-15mm} \times \Biggl[ \Biggl\{ \biggl( \frac{\partial \xi_{\bm{k} A}}{\partial k_{\mu}} \biggr) \hspace{0.7mm} T_{\bm{k} -} T_{\bm{k} +} - \biggl( \frac{\partial \xi_{\bm{k} B}}{\partial k_{\mu}} \biggr) \hspace{0.7mm} \sqrt{1 - T^{2}_{\bm{k} -}} \sqrt{1 - T^{2}_{\bm{k} +}} \Biggr\} D_{14} (\bm{k}) \notag \\
  &\hspace{-15mm} - \Biggl\{ \biggl( \frac{\partial \xi_{\bm{k} A}}{\partial k_{\mu}} \biggr) \hspace{0.7mm} \sqrt{1 - T^{2}_{\bm{k} -}} \sqrt{1 - T^{2}_{\bm{k} +}} - \biggl( \frac{\partial \xi_{\bm{k} B}}{\partial k_{\mu}} \biggr) \hspace{0.7mm} T_{\bm{k} -} T_{\bm{k} +} \Biggr\} D_{23} (\bm{k}) \Biggr] , \label{eq:APPENDIX_03}
\end{align}
and
\begin{equation}
  \sigma^{{\rm SHE}(\rm NC)}_{\mu \mu} = \frac{1}{N} \sum_{\bm{k}} \sigma^{{\rm SHE}(\rm NC)}_{\mu \mu} (\bm{k}), \label{eq:APPENDIX_04} \\
\end{equation}
\begin{align}
  \sigma^{{\rm SHE}(\rm NC)}_{\mu \mu} (\bm{k}) &= \alpha g \mu_{\rm{B}} e \frac{\cos k_{\mu} \sin k_{\bar{\mu}}}{|\bm{g}_{\bm{k}}|} 
\notag \\ 
& \hspace{-15mm} \times \Biggl[ T_{\bm{k} -} \sqrt{1 - T^{2}_{\bm{k} +}} - T_{\bm{k} +} \sqrt{1 - T^{2}_{\bm{k} -}}  \Biggr] \notag \\
  &\hspace{-15mm} \times \Biggl[ \Biggl\{ \biggl( \frac{\partial \xi_{\bm{k} A}}{\partial k_{\mu}} \biggr) \hspace{0.7mm} T_{\bm{k} -} \sqrt{1 - T^{2}_{\bm{k} +}} + \biggl( \frac{\partial \xi_{\bm{k} B}}{\partial k_{\mu}} \biggr) \hspace{0.7mm} T_{\bm{k} +} \sqrt{1 - T^{2}_{\bm{k} -}} \Biggr\} D_{12} (\bm{k}) \notag \\
  &\hspace{-15mm} - \Biggl\{ \biggl( \frac{\partial \xi_{\bm{k} A}}{\partial k_{\mu}} \biggr) \hspace{0.7mm} T_{\bm{k} +} \sqrt{1 - T^{2}_{\bm{k} -}} + \biggl( \frac{\partial \xi_{\bm{k} B}}{\partial k_{\mu}} \biggr) \hspace{0.7mm} T_{\bm{k} -} \sqrt{1 - T^{2}_{\bm{k} +}} \Biggr\} D_{34} (\bm{k}) \Biggr] . \label{eq:APPENDIX_05}
\end{align}
Because $T_{\bm{k} \pm}$ and $D_{a b} (\bm{k})$ are even functions with respect to the wave vector $\bm{k}$, and $\partial \xi_{\bm{k} l}/\partial k_{\mu} \propto \sin k_{\mu}$, 
$\sigma^{{\rm SHE}(\rm C)}_{\mu \mu} (\bm{k})$ and $\sigma^{{\rm SHE}(\rm NC)}_{\mu \mu} (\bm{k})$ satisfy 
\begin{align}
&  \sigma^{{\rm SHE}(\rm C,NC)}_{\mu \mu} (k_{x}, k_{y}) = \sigma^{{\rm SHE}(\rm C,NC)}_{\mu \mu} (-k_{x}, -k_{y}) 
\notag \\ 
=& - \sigma^{{\rm SHE}(\rm C,NC)}_{\mu \mu} (-k_{x}, k_{y}) = - \sigma^{{\rm SHE}(\rm C,NC)}_{\mu \mu} ( k_{x}, -k_{y}). \label{eq:APPENDIX_06}
\end{align}
From Eqs.~(\ref{eq:APPENDIX_02}), (\ref{eq:APPENDIX_04}), and (\ref{eq:APPENDIX_06}), 
the diagonal spin Hall conductivity $\sigma^{\rm{SHE}}_{\mu \mu}$ vanishes after the summation for $\bm{k}$.

Next, we show the analytic form of the diagonal magnetoelectric coefficient in the EO state 
\begin{equation}
  \Upsilon_{\mu \mu} = \Upsilon^{\rm intra}_{\mu \mu} + \Upsilon^{{\rm inter}(\rm C)}_{\mu \mu} + \Upsilon^{{\rm inter}(\rm NC)}_{\mu \mu} , \label{eq:APPENDIX_08}
\end{equation}
where 
\begin{equation}
  \Upsilon^{\rm intra}_{\mu \mu} = \frac{1}{N} \sum_{\bm{k}} \Upsilon^{\rm intra}_{\mu \mu} (\bm{k}) , \label{eq:APPENDIX_09}
\end{equation}
\begin{align}
  &\Upsilon^{\rm intra}_{\mu \mu} (\bm{k}) = \frac{g \mu_{\rm{B}} e}{4 \gamma} \frac{\sin k_{\bar{\mu}}}{|\bm{g}_{\bm{k}}|} \hspace{0.7mm} \Biggl[ \biggl( \frac{\partial E_{\bm{k} 1}}{\partial k_{\mu}} \biggr) \hspace{0.7mm} \{ - f' (E_{\bm{k} 1}) \} \notag \\
  &- \biggl( \frac{\partial E_{\bm{k} 2}}{\partial k_{\mu}} \biggr) \hspace{0.7mm} \{ - f' (E_{\bm{k} 2}) \} \hspace{0mm} + \biggl( \frac{\partial E_{\bm{k} 3}}{\partial k_{\mu}} \biggr) \hspace{0.7mm} \{ - f' (E_{\bm{k} 3}) \} - \biggl( \frac{\partial E_{\bm{k} 4}}{\partial k_{\mu}} \biggr) \hspace{0.7mm} \{ - f' (E_{\bm{k} 4}) \} \Biggr] , \label{eq:APPENDIX_10}
\end{align}
and
\begin{equation}
  \Upsilon^{{\rm inter}(\rm C)}_{\mu \mu} = \frac{1}{N} \sum_{\bm{k}} \Upsilon^{{\rm inter}(\rm C)}_{\mu \mu} (\bm{k}) , \label{eq:APPENDIX_11}
\end{equation}
\begin{align}
  \Upsilon^{{\rm inter}(\rm C)}_{\mu \mu} (\bm{k}) =& - \alpha g \mu_{\rm{B}} e \frac{\cos k_{\mu} \sin k_{\mu} \sin k_{\bar{\mu}}}{|\bm{g}_{\bm{k}}|^{2}} 
\notag \\ 
&\hspace{0.7mm} \times \Biggl[ T^{2}_{\bm{k} -} + T^{2}_{\bm{k} +} - 1 \Biggr] \hspace{0.7mm} \Biggl[ - I_{41} (\bm{k}) + I_{32} (\bm{k}) \Biggr] , \label{eq:APPENDIX_12}
\end{align}
\begin{equation}
  \Upsilon^{{\rm inter}(\rm NC)}_{\mu \mu} =  \frac{1}{N} \sum_{\bm{k}} \Upsilon^{{\rm inter}(\rm NC)}_{\mu \mu} (\bm{k}) , \label{eq:APPENDIX_13}
\end{equation}
\begin{align}
  \Upsilon^{{\rm inter}(\rm NC)}_{\mu \mu} (\bm{k}) =& - \alpha g \mu_{\rm{B}} e \frac{\cos k_{\mu} \sin k_{\mu} \sin k_{\bar{\mu}}}{|\bm{g}_{\bm{k}}|^{2}} 
\notag \\
& \hspace{0.7mm}  \times \Biggl[ T^{2}_{\bm{k} -} - T^{2}_{\bm{k} +} \Biggr] \hspace{0.7mm} \Biggl[ - I_{21} (\bm{k}) + I_{43} (\bm{k}) \Biggr] . \label{eq:APPENDIX_14}
\end{align}
The fact that $\partial E_{\bm{k} a} / \partial k_{\mu} \propto \sin k_{\mu}$ and $I_{a b} (\bm{k})$ is an even function with respect to $\bm{k}$ leads to
\begin{align}
&  \Upsilon^{\rm intra,inter(C),inter(NC)}_{\mu \mu} (k_{x}, k_{y}) = \Upsilon^{\rm intra,inter(C),inter(NC)}_{\mu \mu} (-k_{x}, -k_{y}) 
\notag \\
=& - \Upsilon^{\rm intra,inter(C),inter(NC)}_{\mu \mu} (-k_{x}, k_{y}) = - \Upsilon^{\rm intra,inter(C),inter(NC)}_{\mu \mu} (k_{x}, -k_{y}). \label{eq:APPENDIX_15}
\end{align}
Thus, the diagonal magnetoelectric coefficient $\Upsilon_{\mu \mu}$ disappears after the summation for $\bm{k}$.

\section{Out-of-plane Anisotropy of the Magnetic Susceptibility}
Here we discuss out-of-plane anisotropy of the spin susceptibility. 
First, we show the analytic form of $\chi_{zz}$ and $\Delta \chi_{\perp} \equiv \chi_{zz} - \left( \chi_{xx} + \chi_{yy}\right)/2$. 
By using Eqs.~(\ref{spin_susceptibility_Pauli_and_Van_Vleck})-(\ref{spin_susceptibility_Van_Vleck}), 
we obtain the Pauli part and the Van Vleck part in the normal state 
\begin{align}
  \chi^{\rm{P}}_{zz} &= \frac{g^{2} \mu^{2}_{\rm{B}}}{4 N} \sum_{\bm{k}, a} \{T^{(2)} (\bm{k})\}^{2} \{-f' (E_{\bm{k} a})\}, \label{analytic_form_chi_zz_Pauli_Normal} \\
  \chi^{\rm{VV}}_{zz} &= -\frac{g^{2} \mu^{2}_{\rm{B}}}{2 N} \sum_{\bm{k}} \{T^{(1)} (\bm{k})\}^{2} \{F_{14} (\bm{k}) + F_{23} (\bm{k})\}. \label{analytic_form_chi_zz_Van_Vleck_Normal}
\end{align}
On the other hand, we obtain 
\begin{align}
  \chi^{\rm{P}}_{zz} =& 0, \label{analytic_form_chi_zz_Pauli_EO} \\
  \chi^{\rm{VV}}_{zz} =& -\frac{g^{2} \mu^{2}_{\rm{B}}}{2 N} \sum_{\bm{k}}
  \Bigl[ \{T^{(1)} (\bm{k})\}^{2} \{F_{14} (\bm{k}) + F_{23} (\bm{k})\}
\notag \\
&    + \{T^{(2)} (\bm{k})\}^{2} \{F_{12} (\bm{k}) + F_{34} (\bm{k})\} \Bigr], \label{analytic_form_chi_zz_Van_Vleck_EO}
\end{align}
in the EO state. 

\begin{figure}[htbp]
 \begin{center}
\hspace*{15mm}
   \includegraphics[width=7cm]{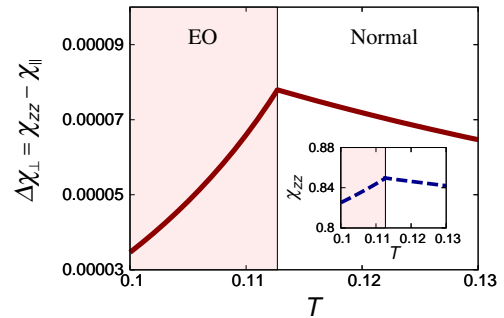}
\vspace{2mm}
   \caption{(Color online) Temperature dependence of out-of-plane anisotropy of the spin susceptibility,
     $\Delta \chi_{\perp} = \chi_{zz} - \chi_{||}$, where $\chi_{||} = (\chi_{xx} + \chi_{yy})/2$.
     The inset shows $\chi_{zz}$. 
     The parameters are the same as Fig.~\ref{torque_and_pseudogap}.
}
   \label{out-of-plane-anisotropy}
 \end{center}
\end{figure}

From Eqs.~(\ref{spin_susceptibility_diagonal_Pauli_EQ})-(\ref{spin_susceptibility_diagonal_Van_Vleck_EO}) and
(\ref{analytic_form_chi_zz_Pauli_Normal})-(\ref{analytic_form_chi_zz_Van_Vleck_EO}),
the analytic form of out-of-plane anisotropy, $\Delta \chi_{\perp} = \Delta \chi^{\rm{P}}_\perp + \Delta \chi^{\rm{VV}}_\perp$, is obtained in the normal state as
\begin{align}
  \Delta \chi^{\rm{P}}_{\perp} &= -\frac{g^{2} \mu^{2}_{\rm{B}}}{8 N} \sum_{\bm{k}, a} \Bigl[ 1 - \{T^{(2)} (\bm{k}) \}^{2} \Bigr] \{-f' (E_{\bm{k} a})\}, \label{analytic_form_chi_perp_Pauli_Normal} \\
  \Delta \chi^{\rm{VV}}_{\perp} &= -\frac{g^{2} \mu^{2}_{\rm{B}}}{4 N} \sum_{\bm{k}} \{T^{(1)} (\bm{k})\}^{2} \{F_{14} (\bm{k}) + F_{23} (\bm{k})\}, \label{analytic_form_chi_perp_Van_Vleck_Normal}
\end{align}
while in the EO state as 
\begin{align}
  \Delta \chi^{\rm{P}}_{\perp} =& -\frac{g^{2} \mu^{2}_{\rm{B}}}{8 N} \sum_{\bm{k}, a} \{-f' (E_{\bm{k} a})\}, \label{analytic_form_chi_perp_Pauli_EO} \\
  \Delta \chi^{\rm{VV}}_{\perp} =&  -\frac{g^{2} \mu^{2}_{\rm{B}}}{4 N} \sum_{\bm{k}}
  \Bigl[ \{T^{(1)} (\bm{k})\}^{2} \{F_{14} (\bm{k}) + F_{23} (\bm{k})\}
\notag \\ 
    &+ \{T^{(2)} (\bm{k})\}^{2} \{F_{12} (\bm{k}) + F_{34} (\bm{k})\} \Bigr]. \label{analytic_form_chi_perp_Van_Vleck_EO}
\end{align}
Combining these formula with the mean field analysis of the forward scattering model, we calculate temperature dependence 
of out-of-plane anisotropy of the spin susceptibility. The obtained result in Fig.~\ref{out-of-plane-anisotropy} is consistent 
with the magnetic torque measurement~\cite{Sato_YBCO}.

\end{document}